\documentclass[11pt]{iopart}
\pdfoutput=1
\usepackage{iopams}
\usepackage{todonotes}

\usepackage{algorithm}
\usepackage{algorithmicx}
\usepackage{algpseudocode}
\algrenewcommand\algorithmicrequire{\textbf{Input:}}
\algrenewcommand\algorithmicensure{\textbf{Output:}}
\algdef{SE}[FOR]{NoDoFor}{EndFor}[1]{\algorithmicfor\ #1}{\algorithmicend\ \algorithmicfor}%

\usepackage{dsfont}
\expandafter\let\csname equation*\endcsname\relax
\expandafter\let\csname endequation*\endcsname\relax
\usepackage{amsmath}
\usepackage{microtype,url}

\usepackage{graphics}
\usepackage{caption}
\usepackage{subcaption}
\usepackage{braket}

\usepackage{citesort}

\usepackage{amsthm}

\usepackage{longtable}

\newcommand{\BigO}[1]{\ensuremath{\operatorname{O}\left(#1\right)}}

\newcommand\defeq{\mathrel{\overset{\makebox[0pt]{\mbox{\normalfont\tiny\sffamily def}}}{=}}}

\usepackage{setspace}

\begin{document}

\title{Accurate and precise characterization of linear optical interferometers}

\author{Ish Dhand$^{1,2,*}$, Abdullah Khalid$^{1}$, He Lu$^{2,3}$, and Barry~C.~Sanders$^{1,2,3,4}$}
\address{$^{1}$Institute for Quantum Science and Technology, University of Calgary, Alberta T2N~1N4, Canada}
\address{$^{2}$Shanghai Branch, National Laboratory for Physical Sciences at Microscale and Department of Modern Physics, University of Science and Technology of China, Shanghai 201315, China}
\address{$^{3}$Synergetic Innovation Center in Quantum Information and Quantum Physics, University of Science and Technology of China, Shanghai 201315, China}
\address{$^{4}$Program in Quantum Information Science, Canadian Institute for Advanced Research, Toronto, Ontario M5G~1Z8, Canada}
\ead{$^{*}$idhand@ucalgary.ca}

\begin{abstract}
We combine single- and two-photon interference procedures for characterizing any multi-port linear optical interferometer accurately and precisely.
Accuracy is achieved by estimating and correcting systematic errors that arise due to spatiotemporal and polarization mode mismatch.
Enhanced accuracy and precision are attained by fitting experimental coincidence data to curve simulated using measured source spectra.
We employ bootstrapping statistics to quantify the resultant degree of precision.
A scattershot approach is devised to effect a reduction in the experimental time required to characterize the interferometer.
The efficacy of our characterization procedure is verified by numerical simulations.
\end{abstract}

\pacs{03.67.Lx, 42.25.Hz, 42.50.Ex, 42.79.Fm}
\maketitle

\section{Introduction}
Linear optics is important in quantum computation and communication.
The simulation of a linear optical interferometer is computationally hard classically subject to reasonable conjectures~\cite{Aaronson2013}.
Single-photon detectors and linear optical interferometers allow for efficient universal quantum computation via linear optical quantum computing (LOQC)~\cite{Knill2001}.
Linear optics can simulate the quantum quincunx~\cite{Do2005} and quantum random walks~\cite{Jeong2004}.
Linear optics coupled with laser-manipulated atomic ensembles enables long-distance quantum communication~\cite{Duan2001}.
A wide class of communication protocols can be realized with coherent states and linear optics~\cite{Arrazola2014}.

Recent advances in photonic technology including photonic circuits on silicon chips~\cite{Politi2008,Matthews2009,Crespi2011,Shadbolt2012,Metcalf2013}, noise-free high-efficiency photon number-resolving detectors~\cite{GolTsman2001,Rosenberg2005,Gansen2007,Lita2008,Najafi2015}, high-fidelity single-photon sources~\cite{Santori2002,URen2004,Faraon2008,Sipahigil2012} have engendered the experimental implementation of multi-port linear optical interferometry.
Reconfigurable linear optical interferometers that can perform arbitrary unitary transformations have been demonstrated~\cite{Mower2014,Harris2015,Carolan2015}.


The accurate and precise characterization of linear optics is important in quantum information processing tasks such as BosonSampling, LOQC and quantum walks.
BosonSampling involves sampling from the output photon coincidence distribution of an interferometer on single photon inputs to each mode.
Sampling from this distribution is computationally hard classically but is easy with a linear-optical interferometer.
The classical hardness of the BosonSampling problem crucially depends on the error in the linear optical interferometer~\cite{Arkhipov2014}.
Similarly, the practical applications of BosonSampling, in quantum metrology and in the computation of molecular vibronic spectra, rely on the accurate implementation and characterization of linear optics~\cite{Motes2015,Huh2014}.

Accurate and precise characterization is important in LOQC because a high success probability of the employed non-deterministic linear-optical gates relies on implementing the desired gates with high fidelity~\cite{Kok2007}. 
Furthermore, linear interferometers used in photonic quantum walks, which display strong non-classical correlations, require accurate characterization especially if quantum walks are employed for solving classically hard problems~\cite{Gamble2010,Peruzzo2010,Sansoni2012}.
In other words, that accurate and precise characterization of interferometers enables a verifiable quantum speedup of linear-optical protocols over classical computers.

Classical-light procedures~\cite{Lobino2008,Keshari2013} for linear optics characterization are unsuitable for Fock-state based experiments because the interferometer parameters change during the coupling and decoupling of classical light sources and of homodyne detectors at the interferometer ports. 
This change could result from drift of interferometer parameters in the time required to couple sources and detectors or as the result of mechanical process of coupling itself.
Characterization procedures that rely on Fock-state (rather than classical-light) inputs are thus more desirable in BosonSampling and LOQC implementations; such procedures would enable interferometer characterization without altering the experimental setup and would thus be accurate.

The Laing-O'Brien procedure~\cite{Laing2012} uses one- and two-photons for characterizing linear optical interferometers and is stable to the length scale of a photon packet. 
This procedure assumes perfect matching in source field and large-number statistics on the detected photons. 
Hence, implementations of this procedure are inaccurate due to spatiotemporal and polarization mode mismatch in the source field and imprecise due to shot noise.

We aim to devise an accurate and precise procedure that uses one- and two-photons for the characterization of linear optical interferometers and to devise a rigorous method to estimate the standard deviation in the linear optical interferometer parameters~\cite{Altepeter2005}.
Furthermore, we aim to provide a correct alternative to the $\chi^2$-test, which has been used to estimate the confidence in the characterized interferometer parameters in current BosonSampling implementations~\cite{Crespi2013,Metcalf2013,Tillmann2014}\footnote{The $\chi^2$-test~\cite{Pearson1900,Plackett1983,Greenwood1996} is used to quantify the goodness of fit between probability distribution functions of two categorical variables, which can take a fixed number of values.
Coincidence-count curves and visibilities are not probability distribution functions of categorical variables, but rather are collections of many categorical variables (variables that can take on one of a fixed finite number of possible values), one variable corresponding to each time-delay value chosen in the experiment.
Hence, quantifying the goodness of fit between two coincidence curves using the $\chi^2$-test is incorrect.
This incorrectness undermines the claim that the data are consistent with quantum predictions and disagree with classical theory~\cite{Metcalf2013,Tillmann2014} and leaves the choice of unitary matrices~\cite{Crespi2013} unjustified.
}.

Here we devise a procedure to characterize a linear optical interferometer accurately and precisely using one- and two-photon interference.
Four strengths of our approach over the Laing-O'Brien procedure~\cite{Laing2012} are that our procedure 
(i)~accounts for and corrects systematic error from spatial and polarization source-field mode mismatch via a calibration procedure (Section~\ref{Sec:Procedure}); 
(ii)~increases accuracy and precision by fitting experimental coincidence data to curve simulated using measured source spectra (Section~\ref{Sec:Procedure});
(iii)~accurately estimates the error bars on the characterized interferometer parameters via a bootstrapping procedure (Section~\ref{Sec:Bootstrapping}); and 
(iv) reduces the experimental time required to characterize interferometers using a scattershot procedure (Section~\ref{Sec:Scattershot}).
%
%
\section{Background}
\label{Sec:Background}
This section provides the background for our one- and two-photon characterization procedure.
The action of a multi-port linear optical interferometer on single photons entering one or two input ports and vacuum entering the other ports is detailed.
Specifically, we calculate the probability of detecting a photon at a given output port when a single photon is incident at a given input port.
The section concludes with expressions for the probability of detecting a coincidence measurement when two controllably delayed photons are incident on the interferometer.

\subsection{Action of a linear optical interferometer}
\label{Subsec:LinearOptics}
In this subsection, we define linear optical interferometers by their action on single photons.
We parameterize the unitary transformation effected by an interferometer and present our treatment of losses and dephasing at the interferometer ports.


Consider a single photon entering the $i$-th mode of an $m$-mode interferometer.
The monochromatic photonic creation and annihilation operators acting on the $i$-th and the $j$-th ports obey the canonical commutation relation\footnote{Two monochromatic photons are distinguishable based on the ports that they occupy and on their respective frequencies $\omega_1$ and $\omega_2$.}
\begin{equation}
\left[a_i(\omega_1),a_j^\dagger(\omega_2)\right] = \delta_{ij}\delta(\omega_1-\omega_2)\mathds{1}
\end{equation}
for positive real frequencies $\omega_{1},\omega_{2}$.
The state of a single photon entering the $i$-th mode is 
\begin{equation}
\ket{1}_i = \int_{-\infty}^{\infty} \mathrm{d}\omega f_{i}(\omega) a_i^\dagger(\omega)\ket{0},
\label{Eq:Single}
\end{equation}
where $f_{i}(\omega)$ is the normalized square integrable {spectral function}, $\ket{0}$ is the $m$-mode vacuum state.
The state of two photons entering modes $i$ and $j\ne i$ of the interferometer is
\begin{equation}
|11\rangle_{ij} = \int_{-\infty}^{\infty} \mathrm{d}\omega_1\int_{-\infty}^{\infty}\mathrm{d}\omega_2\,f_i(\omega_1) f_j(\omega_2) a_i^\dagger(\omega_1)a_j^\dagger(\omega_2)|0\rangle
\label{Eq:twophotonstate}
\end{equation}
with exchange symmetry holding if $f_{i}(\omega) = f_{j}(\omega)$.
One- and two-photon states are transformed into superpositions of one- and of two-photon states respectively under the action of the linear interferometer.

We treat linear interferometers as unitary quantum channels acting on the state of the incoming light.
The interferometer transforms the photonic creation and annihilation operators according to
\begin{equation}
a_{j}^\dagger(\omega) \rightarrow \sum_{i=1}^m V_{ij}(\omega) a_{i}^\dagger(\omega)
\label{Eq:interferometeraction}
\end{equation}
and its complex conjugate, where $V(\omega)$ is the transformation matrix of the interferometer.
Photon-number conservation imposes unitarity
\begin{equation}
V^\dagger(\omega) V(\omega) = \mathds{1}
\end{equation}
of the transformation matrix $V(\omega)$ for all real $\omega$.
In general, the elements $\left\{V_{ij}(\omega)\right\}$ of the transformation matrix depend on the frequency of transmitted light.
We assume that the spectral functions of the incoming light are narrow compared to frequencies over which the entries $\{V_{ij}\}$ change noticeably and thus treat $V$ to be frequency-independent.

If only Fock states are incident at the interferometer and only photon-number-counting detection is performed on the outgoing light, then the measurement outcomes are invariant under phase shifts at each input and output port.
That is, interferometer $\hat{V}=D_1VD_2^\dagger$ produces the same measurement outcome as $V$ for any diagonal unitary matrices $D_1$ and $D_2$.
Mathematically, if $D_1, D_2$ are diagonal unitary matrices, then
\begin{equation}
V\sim\hat{V}\iff \hat{V}= D_1VD_2^\dagger
\end{equation}
is an equivalence relation.
Members of the same equivalence class defined by this equivalence relation produce the same number-counting measurement outcomes on receiving Fock-state inputs.

\begin{figure}[h]
\begin{centering}
\includegraphics[width=0.8\textwidth]{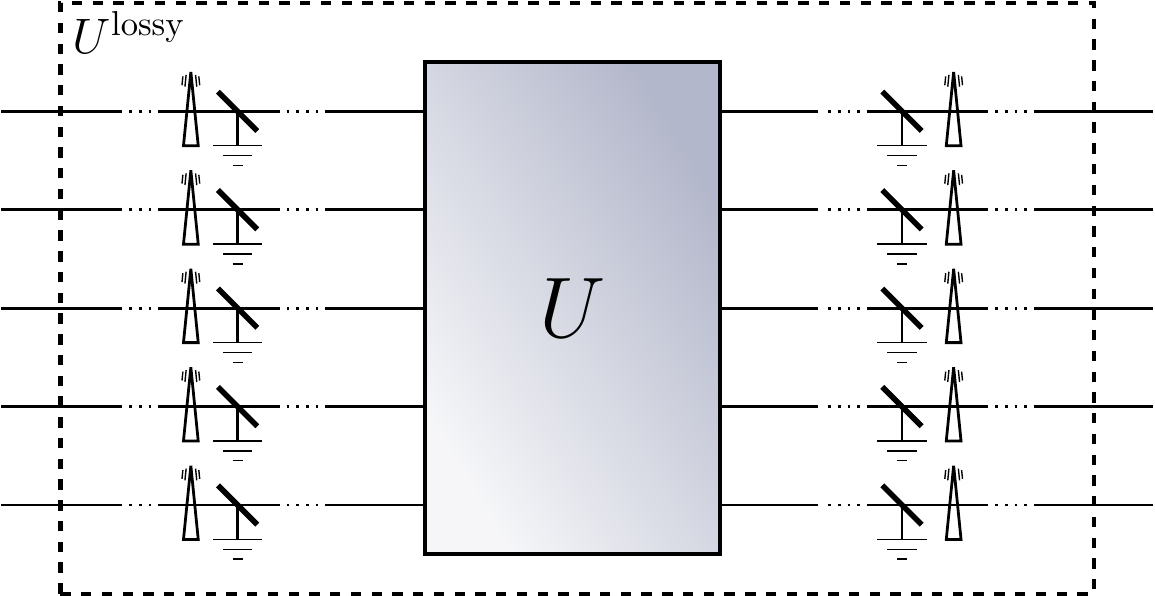}
\caption{Schematic diagram of the interferometer.
$U$ effects a unitary transformation on a multimode state of light.
The dotted lines represent the couplings of the interferometer with light sources and detectors.
The beam splitters at the input and output modes model the linear losses because of imperfect coupling and detector inefficiency.
The vacuum input to these beam splitters is not shown.
One of the beam splitter outputs enters the interferometer while the other one is lost.
The triangles represent the random dephasing at the input and output ports.
The dashed box labelled $U^{\mathrm{lossy}}$ represents the combined effect of the dephasing, the losses and the unitary interferometer.
}
\label{Figure:Interferometer}
\end{centering}
\end{figure}

Each equivalence class can be represented by a unique matrix $U$ whose first row and first column consist of real elements.
The complex matrix entries of the {class representative} $U\sim V$ are 
\begin{equation}
\left\{U_{ij} = t_{ij}\e^{\mathrm{i} \theta_{ij}}:\,
t_{ij}\in \mathds{R}^+,\,
\theta_{ij}\in(-\pi,\pi],\,
\theta_{i1} \equiv 0, \theta_{1j} = 0\,
\forall i,j\in\{1,2,\dots,m\}\right\}.
\label{Eq:Utt}
\end{equation}
The constraints $\theta_{i1} \equiv 0, \theta_{1i} \equiv 0\, \forall i\in\{1,2,\dots,m\}$ on the input and output phases of the transformation matrix are obeyed in the following parameterization of $U$
\begin{align}
U &=
L\times A \times M, \nonumber\\
&\defeq \begin{pmatrix} 1 & 0 & \cdots& 0 \\
0 & \sqrt{\lambda_2} & \cdots & 0\ \\
\vdots & \vdots & \ddots & 0 \\
0 & 0 & \cdots & \sqrt{\lambda_m} \\
\end{pmatrix} \begin{pmatrix}
1 & \cdots & 1 \\
1 & \cdots & \alpha_{2m} \mathrm{e}^{\mathrm{i} \theta_{2m}} \\
\vdots & \ddots & \vdots \\
1 & \cdots & \alpha_{mm} \mathrm{e}^{\mathrm{i} \theta_{mm}}
\end{pmatrix}
\begin{pmatrix} \sqrt{\mu_1} & 0 & \cdots& 0 \\
0 & \sqrt{\mu_2} & \cdots & 0\ \\
\vdots & \vdots & \ddots & 0 \\
0 & 0 & \cdots & \sqrt{\mu_m} \\
\end{pmatrix}. \label{Eq:RMS0}
\end{align}
Thus, the values~$\{\lambda_i\}, \{\alpha_{ij}\},\{\theta_{ij}\}, \{\mu_{j}\}$ completely parametrize the class representative matrix $U$.

The input and output ports of the interferometer are amenable to time-dependent linear loss and dephasing.
We model losses using parameters $\nu_{j}$ and $\kappa_{i}$, which are the respective probabilities of transmission at the input mode $j$ and output mode $i$.
Dephasing is modelled using parameters $\xi_j$ and $\phi_i$, which are the arbitrary multiplicative phases at the input and output ports.
Hence, the actual transformation effected by the interferometer is given by the matrix $U^\mathrm{lossy}$, which has matrix elements
\begin{align}
U^\mathrm{lossy}_{ij} &= \e^{\mathrm{i} \phi_i}\sqrt{\kappa_{i}}\, U_{ij} \sqrt{\nu_{j}}\e^{\mathrm{i} \xi_j}\nonumber\\
&= \e^{\mathrm{i} \phi_i}\sqrt{\kappa_{i}} \sqrt{\lambda_{i}}\, \alpha_{ij}\e^{\mathrm{i} \theta_{ij}} \sqrt{\mu_{j}}\sqrt{\nu_{j}}\e^{\mathrm{i} \xi_j}.
\label{Eq:Probability}
\end{align}
Figure~\ref{Figure:Interferometer} depicts the relation between the representative matrix $U$ and the actual transformation $U^\mathrm{lossy}_{ij}$ that is effected by the interferometer.

This completes our parameterization of the linear optical interferometer.
Our characterization procedure employs one- and two-photon inputs to estimate the values of parameters $\{\lambda_{i}\},\{\alpha_{ij}\},\{\theta_{ij}\},\{\mu_{j}\}$ of~(\ref{Eq:Probability}).
In the next subsection, we recall the expectation values of measurements performed on interferometer outputs when one- and two-photon states are incident at the input ports.

\subsection{One- and two-photon inputs to linear optical interferometer}
\begin{figure}[h]
\includegraphics[width=\textwidth]{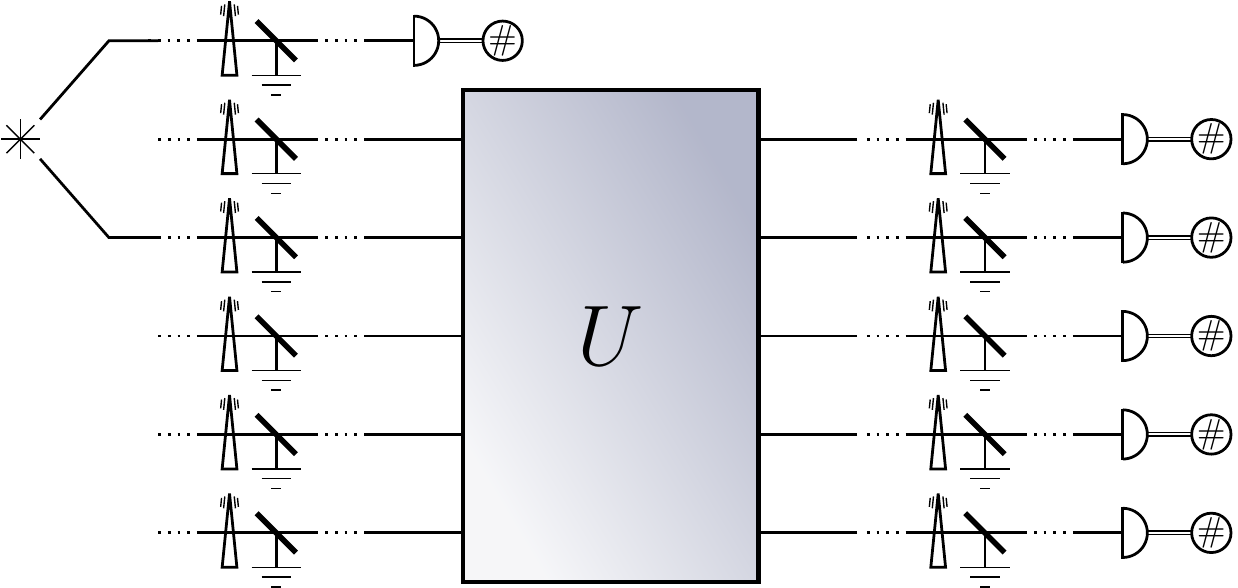}
\caption{Schematic diagram of single-photon counting at the output of an interferometer when single photons are incident at one input port.
The star symbol represents a source of single-photon pairs.
Single photons are incident at one of the input ports while vacuum state is input to the remaining input ports (not shown in figure).
The semicircles at the output ports represent single-photon detectors and the circles with the included $\#$ represent the photon-counting logic connected to the detectors.
}\label{Figure:1Photon}
\end{figure}

Our characterization procedure employs single-photon counting to estimate the amplitudes~$\{\alpha_{ij}\}$ of the representative matrix~$U$ entries.
The arguments~$\{\theta_{ij}\}$ of $U$ are estimated using two-photon coincidence counts.
In this subsection, we give expressions for one- and two-photon transmission probabilities, which are employed in our characterization procedure (Section~\ref{Sec:Procedure}).

We first consider the case of single-photon transmission.
The interferometer transforms the single-photon input state~(\ref{Eq:Single}) to the state at the output ports according to~(\ref{Eq:Probability}).
A photon is detected at the $i$-th output port with a probability
\begin{equation}
P_{ij} = \left|U^\mathrm{lossy}_{ij} \right|^2 = \kappa_{i} \lambda_{i} \alpha_{ij}^2 \mu_{j} \nu_{j}.
 \label{Eq:PSinglePhotons}
\end{equation}
when a single-photon is incident on the $j$-th input port.


Whereas the values of $\{\alpha_{ij}\}$ are estimated using single photon counting, $\{\theta_{ij}\}$ values are estimated using two-photon coincidence measurement.
We now present probabilities of detecting two-photon coincidence at the interferometer outputs when controllably delayed pairs of photon are incident at the input ports.

If a controllably delayed photon pair is incident at input ports $j$ and $j'$, then the probability $C_{ii'jj'}(\tau)$ of coincidence measurement at detectors placed at output ports $i$ and~$i'$ is 
\begin{align}
C_{ii'jj'}(\tau) =&\,
\kappa_{i}\kappa_{i'}\nu_{j}\nu_{j'}\Big[\left(t_{ij}^2t_{i'j'}^2+t_{ij'}^2t_{i'j}^2\right)\int \mathrm{d}\omega_1\mathrm{d}\omega_2 \left|f_{j}(\omega_1)f_{j'}(\omega_2)\right|^2 \nonumber
\\ &+2 t_{ij}t_{ij'}t_{i'j}t_{i'j'} \int \mathrm{d}\omega_1\mathrm{d}\omega_2 f_{j}(\omega_1)f_{j'}(\omega_2)f_{j}(\omega_2)f_{j'}(\omega_1)
\nonumber
\\&\times \cos\left(\omega_2\tau-\omega_1\tau+\theta_{ij}-\theta_{ij'}-\theta_{i'j}+\theta_{i'j'}\right)\Big].
\end{align}
On substituting according to~(\ref{Eq:Utt}), we obtain~\cite{Rohde2006}
\begin{align}
C_{ii'jj'}(\tau) =&\,\kappa_{i}\kappa_{i'}\lambda_{i}\lambda_{i'}\mu_{j}\mu_{j'}\nu_{j}\nu_{j'}\Big[\left(\alpha_{ij}^2 \alpha_{i'j'}^2+ \alpha_{ij'}^2 \alpha_{i'j}^2\right)\int \mathrm{d}\omega_1\mathrm{d}\omega_2 |f_{j}(\omega_1)f_{j'}(\omega_2)|^2 \nonumber
\\ &+2 \gamma \alpha_{ij} \alpha_{ij'} \alpha_{i'j} \alpha_{i'j'} \int \mathrm{d}\omega_1\mathrm{d}\omega_2 f_{j}(\omega_1)f_{j'}(\omega_2)f_{j}(\omega_2)f_{j'}(\omega_1)
\nonumber
\\&\times \cos\left(\omega_2\tau-\omega_1\tau+\theta_{ij}-\theta_{ij'}-\theta_{i'j}+\theta_{i'j'}\right)\Big].
\label{Eq:CoincidenceRate}
\end{align}
 where $\tau$ is the time delay between the two photons, $f_j(\omega), f_{j'}(\omega)$ describe the spectrum of light just before it enters the detectors and $\gamma$ is the mode-matching parameter, which we described in the remainder of this section. 

\begin{figure}[h]
\includegraphics[width=\textwidth]{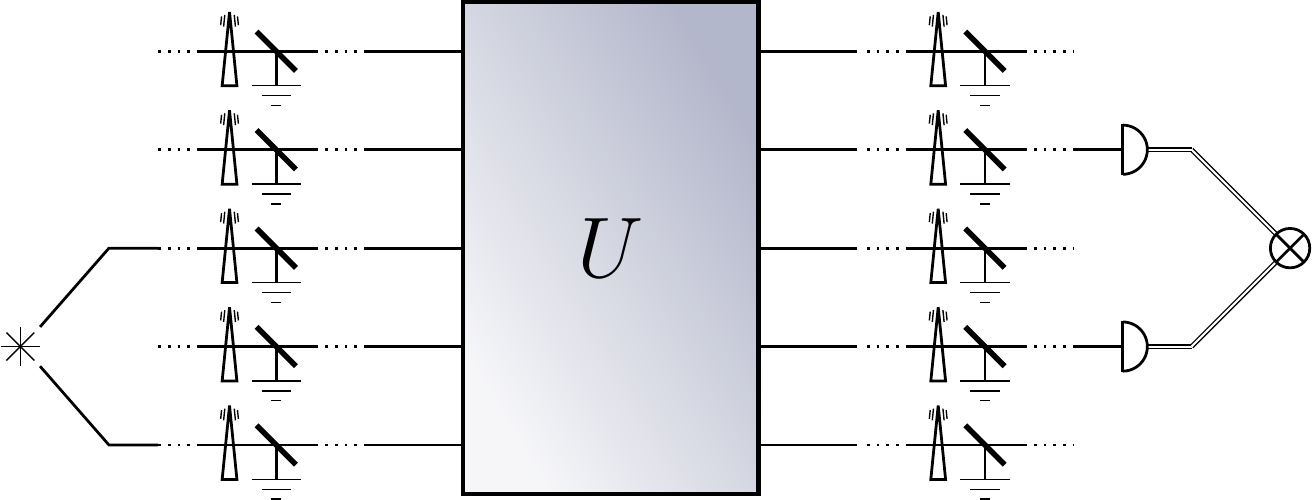}
\caption{Schematic diagram for coincidence measurement the interferometer output when single-photon pairs are incident on two different input ports of an interferometer.
The star symbol represents a source of single-photon pairs and the semicircles at the output ports represent single-photon detectors.
The coincidence logic, which is depicted by $\otimes$, counts two-photon coincidence events at the detectors.
}
\label{Figure:2Photons}
\end{figure}

\begin{figure}[h]
\begin{centering}
\includegraphics[width=0.8\textwidth]{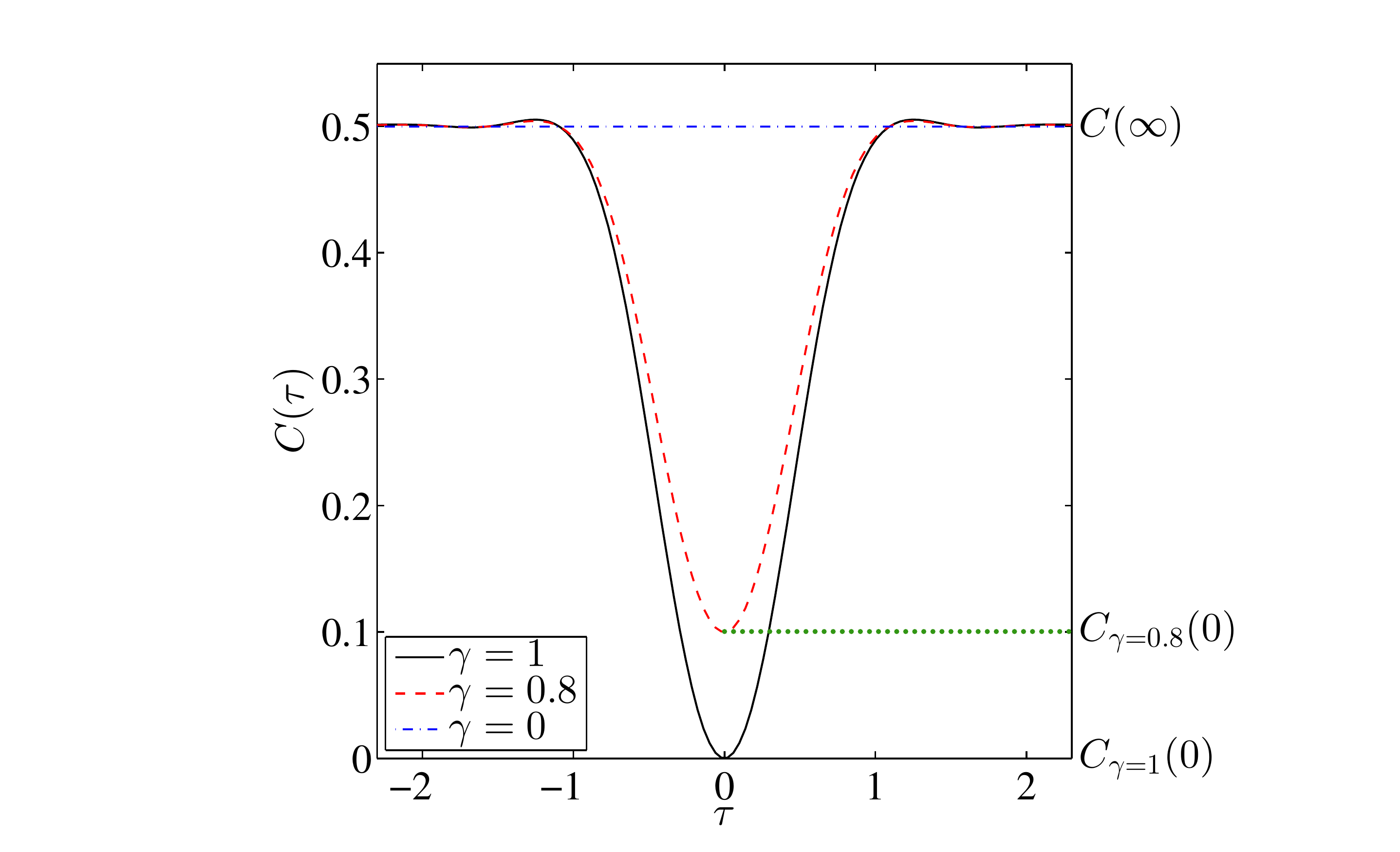}
\caption{
Plots of coincidence probability versus time delay for different values of $\gamma$ for a lossless balanced beam splitter. 
The time delay $\tau$ is in units inverse special width of incoming photons. 
}
\label{Figure:ModeMismatch}
\end{centering}
\end{figure}

Two-photon coincidence probabilities~(\ref{Eq:CoincidenceRate}) depend on the mode matching in the source field.
Spatial and polarization mode mismatch is quantified by the mode-matching parameter $\gamma$~\cite{Rohde2006}.
Perfectly indistinguishable light sources, such as light from a single-mode fibre, have relative mode matching $\gamma=1$ whereas $\gamma=0$ indicates that the sources are completely distinguishable.
Figure~\ref{Figure:ModeMismatch} depicts how imperfect mode matching, i.e., $\gamma<1$, alters the observed two-photon coincidence counts.
Our calibration procedure estimates and accounts for imperfect mode matching, which is assumed to be constant over the runtime of the characterization experiment.


The calculation of the expected coincidence probabilities as a function of the time delay between the photons is detailed in Algorithm~\ref{Alg:Coincidence}.
The next section describes how single-photon transmission probabilities~(\ref{Eq:PSinglePhotons}) and two-photon coincidence probabilities~(\ref{Eq:CoincidenceRate}) are used for characterizing the linear optical interferometer.

\section{Characterization of linear optical interferometer}
\label{Sec:Procedure}
In this section, we describe our procedure to characterize linear optical interferometers.
The outline of this section is as follows.
Subsection~\ref{Subsec:Experiment} describes the experimental data required by our characterization procedure.
This experimental data are processed by various algorithms to determine the transformation matrix~(\ref{Eq:RMS0}).
The algorithm to determine the amplitudes $\{\alpha_{ij}\}$ of the transformation-matrix elements is presented in Subsection~\ref{Subsec:Amplitudes}.
In Subsection~\ref{Subsec:Calibration}, we describe the calibration of the source field by determining the mode-matching parameter
~$\gamma$.
The estimation of $\{\theta_{ij}\}$ using two-photon interference is detailed in Subsection~\ref{Subsec:Phases}.
Maximum-likelihood estimation is employed to find the unitary matrix $U$ that best fits the calculated $\{\alpha_{ij}\}, \{\theta_{ij}\}$ values and serves as the representative matrix~(\ref{Eq:RMS0}).
We discuss the calculation of the best-fit unitary representative matrix in Subsection~\ref{Subsec:MaxLikelyUnitary}. 

\subsection{Experimental procedure and inputs to algorithms}
\label{Subsec:Experiment}
Our characterization procedure relies on measuring 
(i)~the spectral function $f_j$ of the source light, (ii)~single-photon detection counts,
(iii)~two-photon coincidence counts from a beam splitter and~(iv)~two-photon coincidence counts from the interferometer.
The measurement data constitute the inputs to our algorithms, which then yield the representative matrix.
Before presenting the algorithms, we detail the experimental procedure and the inputs received by the algorithm in this subsection.

We characterize the spectral function $f(\omega_i)$ of the incoming light for a discrete set $\Omega=\{\omega_1,\omega_2,\dots,\omega_k\}$ of frequencies.
The integer $k$ of frequencies at which the spectral function is characterized is commonly equal to the ratio of the bandwidth to the frequency step of the characterization device. 
The characterized spectral function $f(\omega_i)$ is used to calculate the coincidence probabilities as detailed in Algorithm~\ref{Alg:Coincidence}.

\begin{algorithm}
	\begin{algorithmic}[1]	
	\Require{\Statex 
	\begin{itemize}
		\item $k, \Omega = \{\omega_1,\omega_2,\dots\omega_{k-1},\omega_k\} \in \left(\mathds{R}^+\right)^{k}$ \Comment{Frequencies at which $f_1,f_2$ are given.}
		\item $f_1,f_2:\Omega \to \mathds{R}^+$\Comment{measured spectra.}
		\item $\ell, T= \{\tau_1,\tau_2,\dots,\tau_\ell\}\in(\mathds{R}\cup 0)^\ell$\Comment{Time delay values.}
		\item $A \leftarrow \{\alpha_{ij},\alpha_{ij'},\alpha_{i\j},\alpha_{i'j'}\}\in \left(\mathds{R}^+\cup\,0\right)^{4}$ \Comment{Amplitudes of $2\times 2$ submatrix of $A$~(\ref{Eq:RMS0}).}
		\item $\Theta \leftarrow \theta_{ij},\theta_{ij'},\theta_{i'j},\theta_{i'j'}\in (-\pi,\pi]$\Comment{Phases of $2\times 2$ submatrix of $A$~(\ref{Eq:RMS0}).}
		\item $\gamma\in[0,1]$ \Comment{Mode-matching parameter
 of photon source.}
		\end{itemize}
	}
	\Ensure{\Statex
	\begin{itemize}
	\item $C:T\to \mathds{R}^+$ \Comment{Two-photon coincidence probabilities correct up to multiplicative factor.}
	\end{itemize}
	}

	\Procedure{Coincidence}{$k,\Omega,f_1,f_2,\ell,T,A,\Theta,\gamma$}
 \For{$\tau$ in $T$}
 \State $C(\tau) \leftarrow\textsc{Integrate}\left[F(A,\Phi, f_1,f_2,\gamma,\omega_i,\omega_j,\tau),\{\omega_a\in \Omega,\omega_b\in\Omega\}\right].$
 \Statex \Comment{Numerically integrate RHS of (\ref{Eq:CoincidenceRate}) over $\omega_i,\omega_j$ with $\kappa_i = \kappa_{i'} = \nu_{j} = \nu_{j'} = 1$.}
 \EndFor
	\State \Return $C$
	\caption{\textsc{Coincidence}: Calculates the expected coincidence rate for two-photon interference for a given $2\times 2$ submatrix of an arbitrary SU$(m)$ transformation.}
	\EndProcedure
	\label{Alg:Coincidence}
	\end{algorithmic}
\end{algorithm}

The amplitudes $\{\alpha_{ij}\}$ are determined by impinging single photons at the interferometer and counting single-photon detections at the outputs.
Single-photon counting is repeated multiple $(B\in\mathds{Z}^+)$ times in order to estimate the precision of the obtained $\{\alpha_{ij}\}$ values.
Specifically, the number 
\begin{equation}
N_{ijb_j}\in \mathds{Z}^{+}:\, i,j\in \{1,\dots ,m\}, b_{j}\in\{1,\dots,B\}
\label{Eq:Nijbdef}
\end{equation}
of single-photon detection events are counted at all $m$ output ports $\{i\}$ for single photons impinged at the $j$-th input ports in the $b_{j}$-th repetition.
The counting is then performed for each of the input ports $j\in\{1,\dots,m\}$ of the interferometer.
Algorithm~\ref{Alg:AmplitudeEstimation} uses $\left\{N_{ijb_j}, b_{j}\in\{1,\dots,B\}\right\}$ values to estimate $\alpha_{ij}$ and the standard deviation of the estimate.
The experimental setup for $\{\alpha_{ij}\}$ measurement is depicted in Figure~\ref{Figure:1Photon}.

Arguments $\{\theta_{ij}\}$ are calculated by fitting curves of measured coincidence counts to curves calculated using measured spectra according to~(\ref{Eq:CoincidenceRate}).
\ref{Sec:CurveFitting} elucidates the inputs and outputs of the curve-fitting procedure, such as the Levenberg-Marquardt algorithm~\cite{Levenberg1944,Marquardt1963}, employed by our algorithms.
Before calculating $\{\theta_{ij}\}$, we calibrate the source field for imperfect mode matching by measuring coincidence counts on a beam splitter of known reflectivity.
Controllably delayed single-photon pairs are incident at the two input ports of the beam splitter and coincidence counting is performed on the light exiting from its two output ports.
Algorithm~\ref{Alg:Calibration} details the estimation of $\gamma$ using coincidence counts $C^{\mathrm{cal}}(\tau)$ for time delay $\tau$ between the incoming photons.

The absolute values and the signs of the arguments $\{\theta_{ij}\in(-\pi,\pi]\}$ are calculated separately.
To estimate the absolute values $\{|\theta_{ij}|\}$ of the arguments, pairs of single photons are incident at two input ports $1$ and~$j\in\{2,\dots,m\}$ and coincidence measurement is performed at two output ports $1$ and~$i\in\{2,\dots,m\}$.
The choice of the input and output ports labelled by index $1$ is arbitrary.
The signs 
\begin{equation}
\operatorname{sgn}\theta_{ij} \defeq \begin{cases}
-1 & \mathrm{if}\, \theta_{ij} < 0, \\
0 & \mathrm{if}\, \theta_{ij} = 0, \\
1 & \mathrm{if}\, \theta_{ij} > 0 \end{cases}
\label{Eq:SignDef}
\end{equation}
of the arguments are estimated using an additional $(m-1)^2$ coincidence measurements.
Algorithm~\ref{Alg:PhaseCalcNChannel} details the choice of input and output ports for estimating $\{\operatorname{sgn}\theta_{ij}\}$.
A schematic diagram of the experimental setup for $\{\theta_{ij}\}$ estimation is presented in Figure~\ref{Figure:2Photons}.

\subsection{Single-photon transmission counts to estimate $\{\alpha_{ij}\}$ (Algorithm~\ref{Alg:AmplitudeEstimation})}
\label{Subsec:Amplitudes}
Now we present our procedure to estimate $\{\alpha_{ij}\}$ values using single-photon counting.
Single-photon transmission probabilities are connected to the amplitudes $\{\alpha_{ij}\}$ according to the relation $P_{ij} = \kappa_{i} \lambda_{i} \alpha_{ij}^2 \mu_{j} \nu_{j}$ (\ref{Eq:PSinglePhotons}).
Although the $\{\alpha_{ij}\}$ values can be calculated from single-photon transmission counts, the factors $\{\lambda_{i}\},\{\mu_{j}\}$ cannot.
The transmission probabilities depend on the products of the factors $\{\lambda_{i}\},\{\mu_{j}\}$ and the loss terms $\{\kappa_{i}\},\{\nu_{j}\}$, so $\{\lambda_{i}\},\{\mu_{j}\}$ cannot be measured without prior knowledge of the losses.
The loss terms are usually unknown and can change between experiments.
Hence, we calculate the values of $\{\alpha_{ij}\}$ from single-photon measurements and choose $\{\lambda_{i}\}$ and~$\{\mu_{j}\}$ such that $U = LAM$ is unitary.

The amplitudes $\{\alpha_{ij}\}$ are determined by estimating transmission probabilities.
The probabilities $P_{11},P_{i1},P_{1j},P_{ij}$ of single-photon detection at output ports $1,i$ when single photons are incident at input ports $1,j$ are expresses in terms of the $\alpha_{ij}$ values according to
\begin{equation}
\frac{P_{11}P_{ij}}{P_{1j}P_{i1}}= \frac{\left|r_1\lambda_1\alpha_{11}\mu_1 s_1\right|^2}{\left| r_1\lambda_1\alpha_{1j} \mu_{j} s_{j}\right|^2}\frac{\left| r_{i}\lambda_{i}\alpha_{ij}\mu_{j} s_{j}\right|^2 }{\left| r_i\lambda_i\alpha_{i1}\mu_1 s_1\right|^2}= \left|\frac{\alpha_{11}\alpha_{ij}}{\alpha_{1j}\alpha_{i1}}\right|^2.
\end{equation}
The probabilities $P_{11},P_{i1},P_{1j},P_{ij}$ are estimated by counting transmitted photons.
The definition~(\ref{Eq:RMS0}) of $\alpha_{ij}$ implies that $\alpha_{11}=\alpha_{i1}=\alpha_{1j}=1$.
Hence, the values of $\alpha_{ij}$ are connected to the single-photon transmission probabilities according to
\begin{equation}
\alpha_{ij} = \sqrt{\frac{P_{11}P_{ij}}{P_{1j}P_{i1}}},
\label{Eq:AlphaMeasure0}
\end{equation}
which is independent of the losses at the input and the output ports.

The transmission probabilities $P_{ij}$ are estimated by counting transmitted photons as follows.
The estimated values of $\{\alpha_{ij}\}$ are random variables that are amenable to random error from under-sampling and experimental imperfections.
Thus, data collection is repeated multiple times.
For accurate estimation of $\alpha_{ij}$ and its standard deviation $\delta\alpha_{ij}$, the number $B$ of repetitions is chosen such that the standard deviation of $\{N_{ijb_j}:b_j\in\{1,\dots,B\}\}$ converges in $B$ for all $i,j\in\{1,\dots,m\}$.
The mean and standard deviation of $\{N_{ijb_j}:b_j\in\{1,\dots,B\}\}$ converge for large enough $B$ if the cumulants of the distribution are finite~\cite{James2006}.

\begin{algorithm}[h]
	\begin{algorithmic}[1]
	\Require{\Statex 
	\begin{itemize}	
		\item $m \in \mathds{Z}^+$, \Comment Number of modes of interferometer. 
		\item $N_{ijb_{j}}: \{1,\dots,m\}\times\{1,\dots,m\} \times \{1,\dots,B\}\to \mathds{Z}^+$ 	
		\Statex\Comment{Single-photon detection counts.}
			\item $B \in \mathds{Z}^+$ \Comment Number of times single-photon counting is repeated .
	\end{itemize}
	}
		
	\Ensure{\Statex
	\begin{itemize}
		\item $\{\tilde{\alpha}_{ij}\} \in \left(\mathds{R}^+\cup 0\right)^{m^2}$\Comment{Estimate of $\{\alpha_{ij}\}$~(\ref{Eq:RMS0}).}
	\end{itemize}
	}

	\Procedure{AmplitudeEstimation}{$m,N_{ijb_{j}},B$}
	\For{$i,j \in \{1,\dots,m\}\times\{1,\dots,m\}$}
	\State $\tilde{\alpha}_{ij} \leftarrow$ {\sc Mean}$\left(\sqrt{{N}_{11b_1}{N}_{ijb_j}/{N}_{1jb_j}{N}_{i1b_1}}: b_1,b_j \in \{1,\dots,B\}\right)$ 
	\EndFor
	\State \Return $\left\{\tilde\alpha_{ij}\right\}$
	\EndProcedure
	\caption{\textsc{AmplitudeEstimation}: Uses single-photon detection counts to calculate the amplitudes of the complex entries of the transformation matrix. $\tilde\bullet$ represents our estimate of $\bullet$.}
	\label{Alg:AmplitudeEstimation}
	\end{algorithmic}
\end{algorithm}

The probabilities $P_{ij}$ are estimated by counting single-photon detection events.
Suppose $N_{ijb_j}$ photons are transmitted from input port $j$ to the detector at output port $i$ when ${N}_{b_j}$ photons are incident and~$b_j\in\{1,\dots,B\}$.
For large enough $B$, the transmission probability converges according to
\begin{equation}
P_{ij} \leftarrow \mathrm{mean}\left\{\frac{N_{ijb_j}}{N_{b_j}}:\, b_j\in \{1,\dots,B\}\right\}.
\end{equation}
Likewise, the amplitudes $\{\alpha_{ij}\}$ are estimated by averaging the single-photon detection counts according to
\begin{align}
\alpha_{ij}
= \sqrt{\frac{P_{11}P_{ij}}{P_{1j}P_{i1}}}
&\leftarrow \mathrm{mean}\left\{ \sqrt{\frac{N_{11b_1}}{N_{b_1}}\frac{N_{ijb_j}}{N_{b_j}}\frac{N_{b_j}}{N_{1jb_j}}\frac{N_{b_1}}{N_{i1b_1}}} :\, b_1,b_j\in \{1,\dots,B\}\right\}\nonumber\\
&= \mathrm{mean}\left\{\sqrt{\frac{N_{11b_1}N_{ijb_j}}{N_{1jb_j}N_{i1b_1}}}:\, b_1,b_j\in \{1,\dots,B\}\right\}.
\label{Eq:AlphaMeasure}
\end{align}
The estimate of $\alpha_{ij}$ relies on single-photon counts measured by impinging photons at the first input port repeatedly (repetition index $b_{1}\in\{1,\dots,B\}$) and independently at the $j$-th input port (with repetitions labelled by a different index $b_{j}\in\{1,\dots,B\}$).

Henceforth, we represent our estimate of any parameter $\bullet$ by $\tilde{\bullet}$.
The estimate $\tilde\alpha_{ij}$ calculated using~(\ref{Eq:AlphaMeasure}) is independent of $N_{b_j}$ and thus resistant to variations in the incident-photon number $N_{b_j}$ over different input modes $j$ and different repetitions $b_j$.
Thus, our estimates~$\{\tilde\alpha_{ij}\}$ are accurate in the realistic case of fluctuating light-source strength and coupling efficiencies.

Finally, the standard deviations $\sigma(\tilde\alpha_{ij})$ of our estimates are calculated according to
\begin{equation}
\sigma(\tilde{\alpha}_{ij})\leftarrow \operatorname{std.~dev.}\left(\sqrt{\frac{N_{11b_1}N_{ijb_j}}{N_{1jb_j}N_{i1b_1}}}:\, b_1,b_j\in \{1,\dots,B\}\right),
\end{equation}
which converges for a large enough $B$.
In line with standard nomenclature, we refer to these standard deviations as error bars. 
Algorithm~\ref{Alg:AmplitudeEstimation} details the estimation of $\{\tilde\alpha_{ij}\}$ and error bars on the obtained estimates.

\subsection{Calibration to estimate mode-matching parameter~$\gamma$ (Algorithm~\ref{Alg:Calibration})}
\label{Subsec:Calibration}

In this subsection, we describe the procedure to calibrate our light sources for imperfect mode matching.
The mode-matching parameter $\gamma$ is estimated using one- and two-photon interference on an arbitrary beam splitter.
First, the reflectivity of the beam splitter is determined using single-photon counting~\cite{Laing2012}.
Next, controllably delayed photon pairs are incident at the beam splitter inputs and coincidence counting is performed on the beam splitter output .
We introduce a curve-fitting procedure to estimate the value of $\gamma$ such that~(\ref{Eq:CoincidenceRate}) best fits the measured coincidence counts.

The beam-splitter reflectivity, which is denoted by $\cos\vartheta$, is estimated as follows.
A beam splitter of reflectivity $\cos \vartheta$ effects the $2\times 2$ transformation
\begin{align}
U_{\mathrm{bs}} &= \begin{pmatrix}
\cos\vartheta&\mathrm{i}\sin\vartheta\\\mathrm{i}\sin\vartheta&\cos\vartheta.
\end{pmatrix}\nonumber\\
&= \begin{pmatrix}1&0\\0&\mathrm{i}
\end{pmatrix}
\begin{pmatrix}1&0\\0&\tan\vartheta
\end{pmatrix}
\begin{pmatrix}1&1\\1&-\cot^2\vartheta
\end{pmatrix}
\begin{pmatrix}\cos\vartheta&0\\0&\sin\vartheta
\end{pmatrix}
\begin{pmatrix}1&0\\0&\mathrm{i}
\end{pmatrix},
\label{Eq:BSMatrix}
\end{align}
which is in the form of~(\ref{Eq:RMS0}) with $\alpha_{22} \defeq \cot^2\vartheta$.
The value of $\alpha_{22}$ is estimated using single-photon counting as described in Algorithm~\ref{Alg:AmplitudeEstimation}.
The estimated beam-splitter reflectivity is 
\begin{equation}
\cos\tilde{\vartheta} = \sqrt{\frac{\alpha_{22}}{1-\alpha_{22}}}.
\label{Eq:FindVartheta}
\end{equation}
The error bar on $\cos\tilde{\vartheta}$ is estimated by repeating the photon counting along the lines of Algorithm~\ref{Alg:AmplitudeEstimation}.

\begin{algorithm}[h]
	\begin{algorithmic}[1]
	\Require{\Statex 
	\begin{itemize}	
		\item $k, \Omega = \{\omega_1,\omega_2,\dots\omega_{k-1},\omega_k\} \in \left(\mathds{R}^+\right)^{k}$ \Comment{Frequencies at which $f_1,f_2$ are given.}
		\item $f_1,f_2:\Omega \to \mathds{R}^+$\Comment{Given spectral functions.}
		\item $\ell, T= \{\tau_1,\tau_2,\dots,\tau_\ell\}\in (\mathds{R}\cup 0)^\ell$\Comment{Time delay values coincidence is measured at.}
		\item $C^\mathrm{cal}:T \to \mathds{R}^+$\Comment{Measured coincidence curve.}
		\item $\vartheta \in(-\pi,\pi]$ \Comment{$\cos\vartheta$ is reflectivity of calibrating beam splitter.} 
	\end{itemize}
	}
	
	\Ensure{\Statex
	\begin{itemize}
		\item $\tilde{\gamma}\in[0,1]$ \Comment{Estimate of mode-matching parameter
 of photon source.}
	\end{itemize}
	}
	
	\Procedure{Calibration}{$k, \Omega, f_1,f_2, \ell, T,C^\mathrm{cal},\vartheta$}
	\State $A \leftarrow \{\cos\vartheta,\sin\vartheta,\sin\vartheta,\cos\vartheta\}$ \Comment{Beamsplitter of reflectivity $R$~(\ref{Eq:BSMatrix})}
	\State $\Phi \leftarrow \{0,\pi/2,\pi/2,0\}$\Comment{Beamsplitter of reflectivity $R$~(\ref{Eq:BSMatrix})}
	\State $C(\tau,\gamma)\defeq \textsc{Coincidence}(\Omega, f_1, f_2,T,A,\Phi,\gamma)$	\Comment{The quantities $\Omega, f_1, f_2,R^\mathrm{cal}$ are given. $\gamma$ is unknown. $\textsc{Coincidence}(\Omega, f_1, f_2,T,R^\mathrm{cal},\gamma)$ depends on $\gamma$ and $\tau$}
	\State \Return $\tilde{\gamma} \leftarrow$ \textsc{Fit}$(C(\tau,\gamma),C^\mathrm{cal}(\tau),1/C^\mathrm{cal}(\tau),\mathrm{InitGuesses})$
	\Comment{Least-squares curve fitting to obtain the value of $\gamma$ that minimizes $\frac{\sum_{\tau \in T}|C^\mathrm{cal}(\tau) - C(\tau,\gamma)|^2}{C^\mathrm{cal}(\tau)}$. 
The argument $1/C^\mathrm{cal}(\tau)$ is the weight function~\cite{Strutz2010} that accounts for experimental noise, which is assumed to be proportional to $\sqrt{C(\tau)}$.
Ignore values of $\tau$ at which $C(\tau) = 1$.
\ref{Sec:CurveFitting} details the choice of initial guesses to the algorithm.}
	\EndProcedure
	\caption{\textsc{Calibration} Calculates the mode-matching parameter $\gamma$ of source-field using a beam splitter of known reflectivity.}
	\label{Alg:Calibration}
	\end{algorithmic}
\end{algorithm}

Next we estimate $\gamma$ using two-photon coincidence counting.
Controllably delayed pairs of photons are incident at the two input ports of the beam splitter.
Coincidence measurement is performed at the output ports for different values of time delay between the two photons.
A curve-fitting algorithm is employed to find the best-fit value of $\gamma$, i.e., the value $\tilde{\gamma}$ that minimizes the squared sum of residues between the measured counts and the coincidence counts expected from~(\ref{Eq:CoincidenceRate}) for the beam splitter matrix~(\ref{Eq:BSMatrix}).
Algorithm~\ref{Alg:Calibration} details the calculations of $\tilde{\gamma}$, which is used to estimate $\{\theta_{ij}\}$ values accurately.

\subsection{Two-photon interference to estimate $\{\theta_{ij}\}$ (Algorithms~\ref{Alg:PhaseCalc2Channel}-\ref{Alg:PhaseCalcNChannel})}
\label{Subsec:Phases}
In this subsection, we describe our procedure to estimate the arguments $\{\theta_{ij}\}$ of the representative matrix $U$~(\ref{Eq:RMS0}).
Our procedure requires the measurement of coincidence counts for $2(m-1)^{2}$ different choices of input and output ports.
Of these measurements, $(m-1)^{2}$ are used to estimate the absolute values $\{|\theta_{ij}|\}$ of the arguments and the remaining $(m-1)^{2}$ are used to estimate the signs $\{\operatorname{sgn}\theta_{ij}\}$.
 
The absolute values $\{|\theta_{ij}|\}$ are estimated as follows.
Single-photon pairs are incident at input ports $1$ and~$j$ and coincidence measurements are performed at output ports $1$ and~$i$ for $i,j\in\{2,\dots,m\}$.
The state~(\ref{Eq:twophotonstate}) of a photon pair is transformed under the action of the $2\times 2$ submatrix
{\setlength{\arraycolsep}{1pt}
\begin{equation}
U_{i1j1} = \begin{pmatrix}\sqrt{\kappa_1}&0\\0&\sqrt{\kappa_{i}}\end{pmatrix}
\begin{pmatrix}\sqrt{\lambda_1}&0\\0&\sqrt{\lambda_{i}}\end{pmatrix}
\begin{pmatrix}1 &1\\1&\alpha_{ij}\e^{\mathrm{i} \theta_{ij}}\end{pmatrix} \begin{pmatrix}\sqrt{\mu_1}&0\\0&\sqrt{\mu_{j}}\end{pmatrix}
\begin{pmatrix}\sqrt{\nu_1}&0\\0&\sqrt{\nu_{j}}\end{pmatrix}
\end{equation}}
of $U$ labelled by the rows $1$ and~$i$ and columns $1$ and~$j$.
The probability of detecting a coincidence at the output ports $1,i$ is
\begin{align}
C_{i1j1}(\tau)
=&
\kappa_{j}\kappa_1\lambda_{j}\lambda_1\nu_{i}\nu_1\mu_{i}\mu_1\Big[\left\{\alpha_{ij}^2+1\right\}\int \mathrm{d}\omega_1\mathrm{d}\omega_2 |f_{j}(\omega_1)f_{1}(\omega_2)|^2 \nonumber
\\ &+2 \gamma \alpha_{ij} \int \mathrm{d}\omega_1\mathrm{d}\omega_2 f_{j}(\omega_1)f_{1}(\omega_2)f_{j}(\omega_2)f_{1}(\omega_1) \cos\left(\omega_2\tau-\omega_1\tau+\theta_{ij}\right)\Big],
\label{Eq:CoincidenceRateGH}
\end{align}
which is obtained by setting $i' = j'=1$ in~(\ref{Eq:CoincidenceRate}).

The measured coincidence counts are used to estimate the value of $|\theta_{ij}|$ as follows.
The shape of the coincidence-versus-$\tau$ curve~(\ref{Eq:CoincidenceRateGH}) depends on the values of $\alpha_{ij}$ and~$\theta_{ij}$.
The shape does not depend on the parameters $\kappa_1,\kappa_i,\lambda_1,\lambda_i,\mu_1,\mu_j,\nu_1,\nu_j$, which lead to a constant multiplicative factor to the coincidence expression.
Furthermore, the shape is unchanged under the transformation $\theta_{ij} \to -\theta_{ij}$ for $\theta_{ij} \in (-\pi,\pi]$ if the spectral functions are identical.
Hence, $|\theta_{ij}|$ can be estimated using the shape of the coincidence function~(\ref{Eq:CoincidenceRateGH}) and the values $\{\tilde\alpha_{ij}\}$ estimated using Algorithm~\ref{Alg:AmplitudeEstimation}.
A curve-fitting algorithm estimates the value $|\tilde{\theta}_{ij}|\in[0,\pi]$ that best fits the measured coincidence counts.
The calculation of $\{|\tilde{\theta}_{ij}|\}$ is detailed in Algorithm~\ref{Alg:PhaseCalc2Channel}.

\begin{algorithm}
	\begin{algorithmic}[1]
	\Require{\Statex 
	\begin{itemize}	

		\item $k, \Omega = \{\omega_1,\omega_2,\dots\omega_{k-1},\omega_k\} \in \left(\mathds{R}^+\right)^{k}$ \Comment{$f_1,f_2$ are measured at frequencies $\Omega$.}
	\item $f_1,f_2:\Omega \to \mathds{R}^+$\Comment{measured spectra.}
		\item $\ell, T= \{\tau_1,\tau_2,\dots,\tau_\ell\}\in (\mathds{R}\cup 0)^\ell$\Comment{Time delay values coincidence is measured at.}
	\item $C^{\mathrm{exp}}:T \to \mathds{R}^+$\Comment{Measured coincidence curve.}
 \item $A \leftarrow \{\alpha_{ij},\alpha_{ij'},\alpha_{i'j},\alpha_{i'j'}\}$ \Comment{Complex amplitudes of $2\times 2$ submatrix of $A$~(\ref{Eq:RMS0}).}
	\item $\Theta \leftarrow \{\theta_{ij'},\theta_{i'j},\theta_{i'j'}\in (-\pi,\pi]\}$\Comment{Three complex arguments of submatrix.}
	\item $\gamma$ \Comment{Mode-matching parameter of photon source.}
	\end{itemize}
	}
	
	\Ensure{\Statex
	\begin{itemize}
	\item $|\tilde{\theta}_{ij}|$\Comment{Estimated magnitude of the unknown complex argument.}
	\end{itemize}
	}
	
	\Procedure{Argument2Port}{$k, \Omega,f_1,f_2,\ell, T,C_\mathrm{exp},A,\Theta,\gamma$}
	\State $\Phi \defeq \{\theta_{ij},\theta_{ij'},\theta_{i'j},\theta_{i'j'}\}$\Comment{Set of three known phases and one unknown phase.}
	\State $C(\tau,\theta_{ij})\defeq \textsc{Coincidence}(\Omega, f_1, f_2,T,A,\Phi,\gamma)$
	\State \Return $\tilde{\theta}_{ij} \leftarrow \left|\textsc{LM}(C(\tau,\theta_{ij}),C^{\mathrm{exp}}(\tau),1/C^\mathrm{exp}(\tau))\right|$
	\Statex\Comment{Use curve fitting to compute the $\theta_{ij}$ value that minimizes $\frac{\sum_{\tau \in T}|C_\mathrm{exp}(\tau) - C(\tau,\gamma)|^2}{C_\mathrm{exp}(\tau)}$.}
	\EndProcedure
	\caption{\textsc{Argument2Port}: Calculates the unknown complex argument in the entries of a $2\times 2$ transformation using a two-photon coincidence curve.}
	\label{Alg:PhaseCalc2Channel}
	\end{algorithmic}
\end{algorithm}

Our procedure computes the signs by using an additional $(m-1)^{2}$ coincidence measurements.
First we arbitrarily set $\theta_{22}$ as positive
\begin{equation}
\operatorname{sgn}\theta_{22} = 1
\label{Eq:Theta22}
\end{equation}
 because of the invariance\footnote{
Expectation values of Fock-state projection measurement with Fock-state inputs are unchanged under $U\to U^*$ if the spectral functions are equal $f_1(\omega) = f_2(\omega)$.
Otherwise, the sign of $-\alpha_{22}$ can be ascertained using the difference in the $\tau>0$ and~$\tau<0$ coincidence counts in $C_{2,2,1,1}(\tau)$.

} of one- and two-photon statistics under complex conjugation $U\to U^*$~\cite{Laing2012}.
The signs of the remaining arguments $\{\theta_{ij}\}$ are set using the coincidence counts between output ports $\{i,i'\}$ when photon pairs are incident at input ports $\{j,j'\}$ for a suitable choice of $\{i',j'\}$ as we describe below.
The coincidence probability at the output ports $i,i'$ is 
\begin{align}
C_{ii'jj'}(\tau) =&\,\kappa_{i}\kappa_{i'}\lambda_{i}\lambda_{i'}\mu_{j}\mu_{j'}\nu_{j}\nu_{j'}\Big[\left(\alpha_{ij}^2 \alpha_{i'j'}^2+ \alpha_{ij'}^2 \alpha_{i'j}^2\right)\int \mathrm{d}\omega_1\mathrm{d}\omega_2 |f_{j}(\omega_1)f_{j'}(\omega_2)|^2 \nonumber
\\ &+2 \gamma \alpha_{ij} \alpha_{ij'} \alpha_{i'j} \alpha_{i'j'} \int \mathrm{d}\omega_1\mathrm{d}\omega_2 f_{j}(\omega_1)f_{j'}(\omega_2)f_{j}(\omega_2)f_{j'}(\omega_1)
\nonumber\\
&\times \cos\left(\omega_2\tau-\omega_1\tau+\beta_{ii'jj'}\right)\Big],\label{Eq:Coiniijj}
\end{align}
where
\begin{equation}
\beta_{ii'jj'} \defeq |\theta_{i'j'}-\theta_{ij'}-\theta_{i'j}+\theta_{ij}|\in[0,\pi]\label{Eq:Beta}.
\end{equation}
Curve fitting is employed to estimate the value of $\beta_{ii'jj'}$ that best fits the measured coincidence counts.

\begin{algorithm}[h]
	\begin{algorithmic}[1]
	\Require{\Statex 
	\begin{itemize}	
		\item $\beta \equiv |\theta_{i'j'}-\theta_{ij'}-\theta_{i'j}+\theta_{ij}|$ \Comment As defined in~(\ref{Eq:SignEquation}).
		\item $\theta_{i'j'}, \theta_{ij'}, \theta_{i'j}, \left|\theta_{ij}\right|$ \Comment Equations~(\ref{Eq:SignEquation}-\ref{Eq:BetaMinus}).
	\end{itemize}
	}
		
	\Ensure{\Statex
	\begin{itemize}
		\item $\operatorname{sgn}\theta_{ij}$\Comment{Sign of $\theta\in(-\pi,\pi]$ is defined in~(\ref{Eq:SignDef})}
	\end{itemize}}
		\Procedure{SignCalc}{$\beta,\theta_{i'j'}, \theta_{ij'}, \theta_{i'j}, \left|\theta_{ij}\right|$}
		\State $\beta^+ \leftarrow |\theta_{i'j'}-\theta_{ij'}-\theta_{i'j}+|\theta_{ij}||$ \Comment If $\theta_{gh}<0$, then $\beta = \beta^{-}$.
		\State $\beta^- \leftarrow |\theta_{i'j'}-\theta_{ij'}-\theta_{i'j}-|\theta_{ij}||$ \Comment If $\theta_{gh}>0$, then $\beta = \beta^{+}$.
		\State $\text{{sgn}}\,\theta_{ij} \leftarrow \operatorname{sgn}\left|\beta-\beta^-\right|-\left|\beta-\beta^+\right|$
 		\State\Return $\text{sgn}\,\theta_{ij}$
		\EndProcedure
		
	\caption{\textsc{SignCalc}: Calculates the complex-phase sign of an element of the $2\times 2$ submatrix of an interferometer transformation matrix.}
	\label{Alg:SignCalc}
	\end{algorithmic}
\end{algorithm}

The estimated value of $\beta_{ii'jj'}$ is employed by Algorithm~\ref{Alg:SignCalc} to ascertain the sign of $\theta_{ij}$.
Algorithm~\ref{Alg:SignCalc} relies on the identity
\begin{equation}
\operatorname{sgn}\theta_{ij} = \operatorname{sgn}\left(|\beta_{ii'jj'}-\beta^{-}_{ii'jj'}|-|\beta_{ii'jj'}-\beta^{+}_{ii'jj'}|\right),\label{Eq:SignEquation}
\end{equation}
and on known values of 
\begin{equation}
\beta^{\pm}_{ii'jj'}\defeq |\theta_{i'j'}-\theta_{ij'}-\theta_{i'j} \pm |\theta_{ij}||, \beta^{\pm}_{ii'jj'}\in[0,\pi]\label{Eq:BetaMinus}
\end{equation}
to ascertain the sign of $\theta_{ij}$.
If the sign of $\theta_{ij}$ is positive, then $\beta_{ii'jj'} = \beta_{ii'jj'}^{+}$ and~(\ref{Eq:SignEquation}) returns a positive $\operatorname{sgn}\theta_{ij}$. Otherwise, $\beta_{ii'jj'}=\beta_{ii'jj'}^-$, in which case~(\ref{Eq:SignEquation}) gives a negative sign.

\begin{figure}
\begin{centering}
 \begin{subfigure}{0.24\textwidth}
 \includegraphics[width=\textwidth]{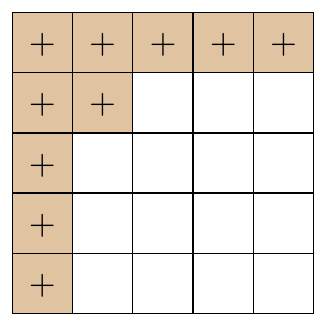}
 \caption{}
 \end{subfigure}
 \begin{subfigure}{0.24\textwidth}
 \includegraphics[width=\textwidth]{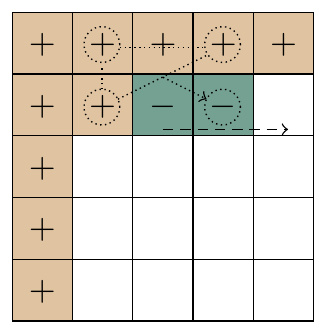}
 \caption{} 
 \end{subfigure}
 \begin{subfigure}{0.24\textwidth}
 \includegraphics[width=\textwidth]{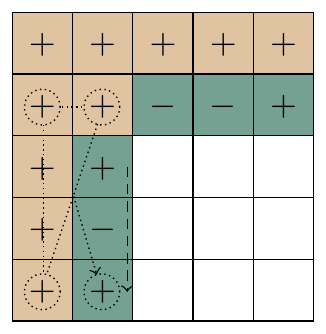}
 \caption{}
 \end{subfigure}
 \begin{subfigure}{0.24\textwidth}
 \includegraphics[width=\textwidth]{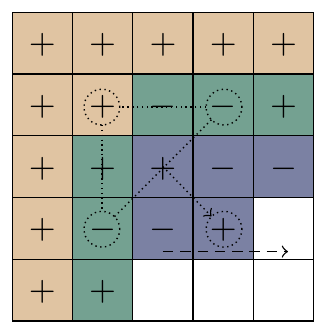}
 \caption{}
 \end{subfigure}
\end{centering}
\caption{%
A depiction of the sign estimation procedure in Lines~\ref{Algline:StartSigns}--\ref{Algline:EndSigns} of Algorithm~\ref{Alg:PhaseCalcNChannel}.
(a)~The first row and first column arguments $\{\theta_{i1}\},\{\theta_{1j}\}$ are zero, so their signs are arbitrarily set as positive. $\theta_{22}$ is set as positive according to~(\ref{Eq:Theta22}).
(b)~The sign of each second row argument $\theta_{i2}$ is set using the known values $|\theta_{22}|,|\theta_{i2}|$ and coincidence measurement for input ports $1,2$ and output ports $2,i$ as in Line~\ref{Alglin:Rows}. 
(c)~The sign of each second column argument $\theta_{2j}$ is set using the known values $|\theta_{22}|,|\theta_{2j}|$ and coincidence measurement for input ports $2,j$ and output ports $1,2$ as in Line~\ref{Alglin:Columns} of Algorithm~\ref{Alg:PhaseCalcNChannel}. 
(d)~The signs of each remaining argument $\theta_{ij}$ is set using the known values $|\theta_{22}|,|\theta_{i2}|,|\theta_{2j}|$ and coincidence measurement for input ports $2,j$ and output ports $2,i$ in Line~\ref{Alglin:Rest} of Algorithm~\ref{Alg:PhaseCalcNChannel}. 
}\label{Figure:Ordering}
\end{figure}

Algorithm~\ref{Alg:PhaseCalcNChannel} iteratively chooses indices $i,i',j,j'$ such that the signs of $\theta_{ij'}, \theta_{i'j},\theta_{i'j'}$ have already been ascertained before ascertaining the sign of $\theta_{ij}$.
In each iteration, the values of $\beta_{ii'jj'}^{\pm}$ are calculated by substituting $|\theta_{ij}|,|\theta_{ij'}|,|\theta_{i'j}|,|\theta_{i'j'}|,\operatorname{sgn}\theta_{ij'},\operatorname{sgn}\theta_{i'j},\operatorname{sgn}\theta_{i'j'}$.
The algorithm estimates $\beta_{ii'jj'}$ by curve fitting measured coincidence counts to~(\ref{Eq:Coiniijj}).
Algorithm~\ref{Alg:SignCalc} is ascertains the sign of $\theta_{ij}$ using the estimates of $\beta_{ii'jj'}$ and $\beta_{ii'jj'}^{\pm}$.
One suitable ordering of indices $ii'jj'$, which we depict in Figure~\ref{Figure:Ordering}, is
\begin{itemize}
\item set $i'=2,j'=1$ to determine sgn$\theta_{i2}$ for $i\in \{3,\dots,m\}$ (Figure~\ref{Figure:Ordering}b), 
\item set $i'=1,j'=2$ to determine sgn$\theta_{2j}$ for $j\in \{3,\dots,m\}$ (Figure~\ref{Figure:Ordering}c),
\item set $i'=2,j'=2$ to determine sgn$\theta_{ij}$ for $(i,j)\in\{3,\dots,m\}\times\{3,\dots,m\}$ (Figure~\ref{Figure:Ordering}d).
\end{itemize}
In summary, $\operatorname{sgn}\theta_{ij}$ is determined using the values of $\beta_{ii'jj'}$, which are estimated by curve fitting, and of~$\beta_{ii'jj'}^\pm$, which are computed using the signs and amplitudes of $\theta_{ij'},\theta_{i'j},\theta_{i'j'}$.
Algorithms~\ref{Alg:PhaseCalc2Channel}-\ref{Alg:PhaseCalcNChannel} detail the step-by-step procedure to determine the absolute values and the signs of $\{\theta_{ij}\}$.

For certain interferometers $U$, the ordering of indices $ii'jj'$ depicted in Figure~\ref{Figure:Ordering} can lead to instability in the characterization procedure. 
\ref{Sec:Instability} elucidates on this instability and presents strategies to counter the instability.
This completes our procedure to characterize the matrix $A$ for representative matrix $U = LAM$.
In the next subsection, we present a procedure to estimate the matrix that is most likely for the characterized matrix $A$.

\begin{algorithm}[h]
	\begin{algorithmic}[1]
	\Require{\Statex 
	\begin{itemize}	
		\item $k, \Omega = \{\omega_1,\omega_2,\dots\omega_{k-1},\omega_k\} \in \left(\mathds{R}^+\right)^{k}$ \Comment{$f_1,f_2$ are measured at frequencies $\Omega$.}
	\item $f_1,f_2:\Omega \to \mathds{R}^+$\Comment{measured spectra.}
		\item $\ell, T= \{\tau_1,\tau_2,\dots,\tau_\ell\}\in (\mathds{R}\cup 0)^\ell$\Comment{Time delay values coincidence is measured at.}
	\item $C^{\mathrm{exp}}_{ii'jj'}(\tau)$ for $(i,i',j,j')\in\{1,2\}\times\{1,\dots,m\}\times\{1,2\}\times\{1,\dots,m\},\,i\ne i', j\ne j'$ 
	\Statex \Comment{Measured coincidence at output ports $i',j'$ when photons that have mutual delay $\tau$ are incident at input ports $i,j$.}
	\item $\tilde\alpha:\{2,\dots,m\}\times\{2,\dots,m\}\to\mathds{R}^+$ \Comment{Complex amplitudes~(\ref{Eq:RMS0}).}
	\item $\gamma\in[0,1]$ \Comment{Mode-matching parameter estimated using Algorithm~\ref{Alg:Calibration}.}
	\end{itemize}
	}	
	\Ensure{\Statex
	\begin{itemize}
		\item $\tilde\theta_{ij}:\{1,\dots,m\}\times\{1,\dots,m\}\to (-\pi,\pi]$\Comment{Complex Arguments~(\ref{Eq:RMS0}).}
	\end{itemize}
	}
	\Procedure{ArgumentCalc}{$k, \Omega,f_1,f_2,\ell, T,C^{\mathrm{exp}}_{ii'jj'}(\tau),\alpha_{ij},\gamma$}
	\For{$i$ in $\{1, \dots, m\}$} 
		\State $\theta_{i1}, \theta_{1i},\,\operatorname{sgn}\theta_{i1},\,\operatorname{sgn}\theta_{1i} \leftarrow 0$ \Comment The first row, column are real valued.
	\EndFor	

	\For{$(i,j)$ in $\{2, \dots, m\}\times\{2, \dots, m\}$}
		\State $A \leftarrow \{1,1,1,\alpha_{gh}\}$, $\Phi \leftarrow \{0,0,0\}$ \Comment $2\times2$ matrix: rows $1,i$, columns $1,j$.
		\State $|\tilde\theta_{ij}| \leftarrow \textsc{Argument2Port}\left(C^{\mathrm{exp}}_{1i1j}T,\Omega, f_1, f_2,T,A,\Phi,\gamma\right)$
	\EndFor\label{Alglin:EndAmplitudes}
	
	\State $\operatorname{sgn}\theta_{22}\leftarrow 1$ \Comment The sign of $\theta_{22}$ is positive by definition.\label{Algline:StartSigns}
	
	\For{($i,i',j,j')\in \{2\}\times\{3,\dots,m\}\times \{2\}\times\{3,\dots,m\}
	\cup \{1\}\times\{2\}\times \{2\}\times\{3,\dots,m\}
	\cup \{2\}\times\{3,\dots,m\}\times \{1\}\times\{2\}$}
	\State $A \leftarrow \{0,0,0\}, \Phi \leftarrow \{0,0,0\}$
	\State $\beta_{i,i',j,j'} \leftarrow \textsc{Argument2Port}(C^{\mathrm{exp}}_{ii'jj'}(\tau),\Omega, f_1, f_2,T,A,\Phi,\gamma)$
	\EndFor
	\For{$i$ in $\{3, \dots, m\}$} \label{Alglin:StartSecondRowColumn}
		\State $\tilde\theta_{i2} \leftarrow |\tilde\theta_{i2}|${\sc SignCalc}$(\beta_{122i},0,\theta_{22},0,|\theta_{i2}|,));$\label{Alglin:Rows}
		\State $\tilde\theta_{2i} \leftarrow |\tilde\theta_{2i}|${\sc SignCalc}$(\beta_{2i12},0,\theta_{22},0,|\theta_{2i}|,));$\label{Alglin:Columns}
	\EndFor\label{Alglin:EndSecondRowColumn}
	
	\For{$(i,j)$ in $\{3, \dots, m\}\times\{3, \dots, m\}$}
		\State $\tilde\theta_{ij} \leftarrow |\tilde\theta_{ij}|${\sc SignCalc}$(\beta_{ii'jj'},\theta_{22},\theta_{i2},\theta_{2j},|\theta_{ij}|)$\label{Alglin:Rest} 
	\EndFor
	\State \Return $\{\theta_{ij}\}$
	\EndProcedure \label{Algline:EndSigns}

	\caption{\textsc{ArgumentCalc}: Calculate $\{\theta_{ij}\}$ using two-photon coincidences}
	\label{Alg:PhaseCalcNChannel}
	\end{algorithmic}
\end{algorithm}


\subsection{Maximum-likelihood estimation for finding unitary matrix}
\label{Subsec:MaxLikelyUnitary}
At this stage, we have estimated the matrix $A$~(\ref{Eq:RMS0}).
The diagonal matrices $L$ and~$M$ can be uniquely determined from $A$ as follows.
The representative matrix $U = LAM$ is unitary so we have
\begin{equation}
U U^\dagger =\mathds{1},
\end{equation}
which, upon substitution $U=LAM$, implies that
\begin{align}
LAMM^*A^\dagger L^* &= \mathds{1}\nonumber\\
\implies AMM^* A^\dagger &= L^{-1}L^{*-1}.\label{Eq:MRRM}
\end{align}
Considering the first columns of the matrices~(\ref{Eq:MRRM}) gives
\begin{equation}
A_{ij}M^*_{jj}M_{jj}A^\dagger_{j1} = \begin{pmatrix}1\\ 0 \\ \vdots\\0\end{pmatrix}
\end{equation}or
\begin{equation}
A\begin{pmatrix}
\mu_1\\\mu_2\\\vdots \\\mu_m\end{pmatrix}
= \begin{pmatrix}1 \\0 \\\vdots \\0\end{pmatrix}.
\label{Eq:MS}
\end{equation}
Similarly, using $U^\dagger U= \mathds{1}$ we obtain
\begin{equation}A^\dagger\begin{pmatrix}1\\\lambda_2\\\vdots \\\lambda_m\end{pmatrix}
=\frac{1}{\mu_1}
\begin{pmatrix}1 \\0 \\\vdots \\0\end{pmatrix}.
\label{Eq:MR}
\end{equation}
Equations~(\ref{Eq:MS}) and~(\ref{Eq:MR}) are systems of linear equations that can be solved for $L$ and~$M$ respectively using standard methods~\cite{Kailath1980}.
The solutions $L$ and~$M$ of the linear systems and the characterized matrix $A$ give us the representative matrix $U = LAM$.

\begin{algorithm}[h]
	\begin{algorithmic}[1]
	\Require{\Statex 
	\begin{itemize}
		\item $\tilde\alpha:\{1,\dots,m\}\times\{1,\dots,m\}\to\mathds{R}^+\cup 0$\Comment{Estimated amplitudes of $A$~(\ref{Eq:RMS0})}.
		\item $\tilde\theta:\{1,\dots,m\}\times\{1,\dots,m\}\to(-\pi,\pi]$\Comment{Estimated arguments of $A$~(\ref{Eq:RMS0})}.
	\end{itemize}
	}
	\Ensure{\Statex
	\begin{itemize}
		\item $W\in \mathrm{SU}(n)$\Comment{Unitary matrix with maximum likelihood of generating $A$.}
	\end{itemize}
	}
	
	\Procedure{MaxLikelyUnitary}{$\alpha_{ij},\theta_{ij}$}
	\State $\lambda_1\leftarrow 1$
	\State $\{\tilde\mu_i: i\in \{1,\dots,m\}\} \leftarrow$ solution of system~(\ref{Eq:MS}) of linear equations.
	\State $\{\tilde\lambda_i: i\in \{2,\dots,m\}\}\leftarrow$ solution of system~(\ref{Eq:MR}) of linear equations.
	\State $\tilde{U}_{ij}\leftarrow \tilde\lambda_{i}\tilde\alpha_{ij}\e^{\mathrm{i}\tilde\theta_{ij}}\tilde\mu_{j} $
	\State $W \leftarrow \left(\tilde{U}\tilde{U}^\dagger\right)^{-\frac{1}{2}}\tilde{U}$\Comment{Assumption: $U_{ij}-\tilde{U}_{ij}$ is an iid Gaussian random variable with zero mean for all $i,j \in\{1,\dots,m\}$.}
	\EndProcedure
	\caption{\textsc{MaxLikelyUnitary}: Calculates unitary matrix that has maximum likelihood of generating estimated $\{\alpha_{ij}\},\{\theta_{ij}\}$}
	\label{Alg:MaxLikelyUnitary}
	\end{algorithmic}
\end{algorithm}

The experimentally determined ${\tilde{A}}$ is different from the actual~$A$ because of random and systematic error in the experiment, where we denote the experimentally determined values of interferometer parameter $\bullet$ by $\tilde{\bullet}$.
Similarly, the $\tilde{L}$ and~$\tilde{M}$ matrices obtained by solving Eqs.~(\ref{Eq:MS}) and~(\ref{Eq:MR}) for $\tilde{A}$ (rather than $A$) differ from the actual $L$ and $M$ respectively.
The estimated $\tilde{U} = \tilde{L}\tilde{A}\tilde{M}$ is thus a non-unitary matrix and is not equal to $U$ in general.
Furthermore, $\tilde{U}$ is a random matrix, which depends on the random errors in the one- and two-photon experimental data.

\begin{figure}[h]
\begin{centering}
\includegraphics[width=0.5\textwidth]{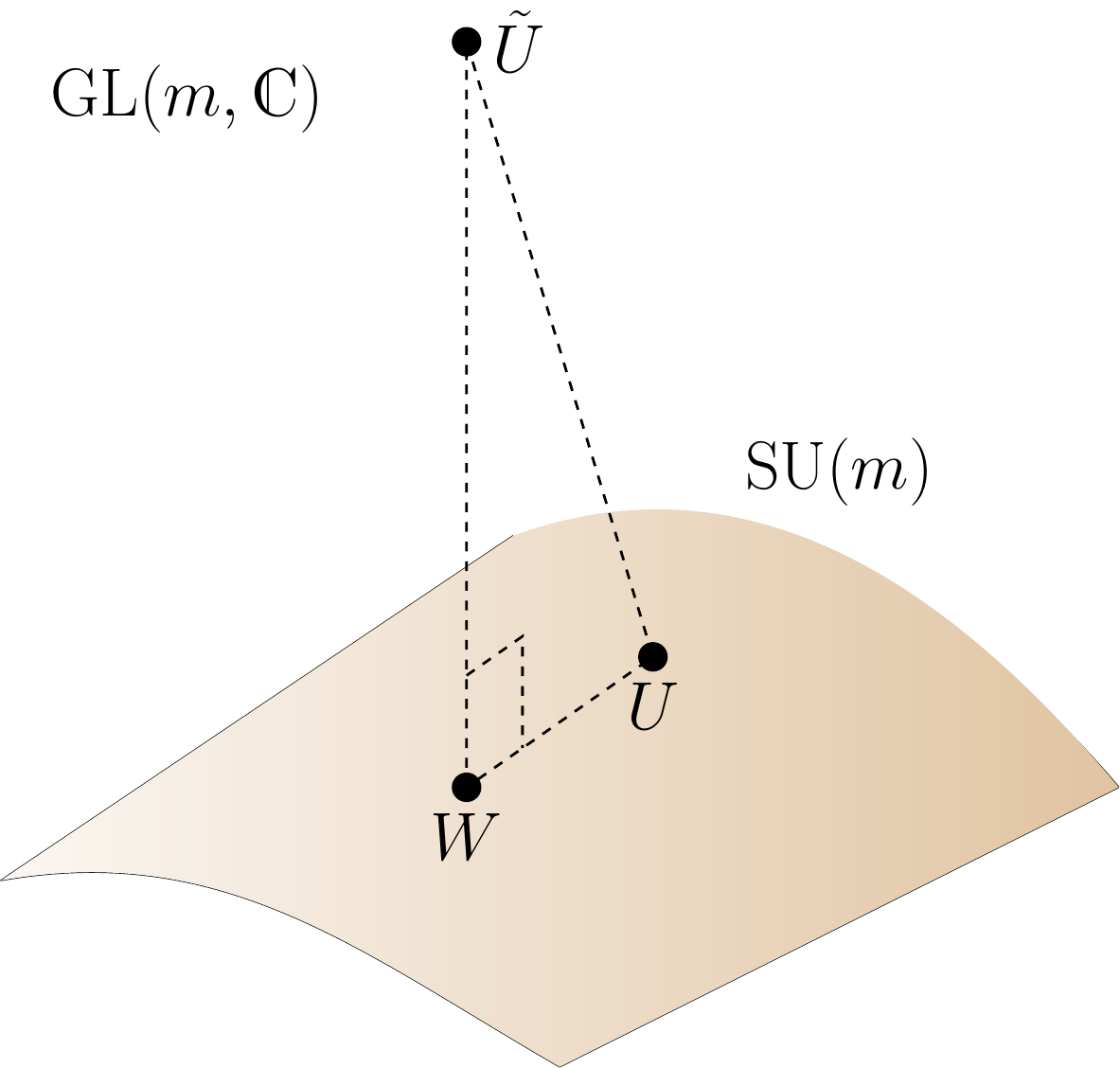}
\caption{
A depiction of the error in reconstruction of the interferometer matrix $U$.
The matrix $U$ represents the unitary transformation effected by the interferometer.
$\tilde U$ is the complex-valued transformation matrix returned by the reconstruction procedure.
Algorithm~\ref{Alg:MaxLikelyUnitary} returns $W$, which represents the unitary matrix that is most likely to have generated the data collected in the characterization experiment.
}
\label{Figure:UUV}
\end{centering}
\end{figure}

We employ maximum-likelihood estimation to calculate the unitary matrix $W$ that best fits the collected data.
First, bootstrapping techniques are used to estimate the probability-density function (pdf) of the entries of the random matrix $\tilde{U}$~\cite{Efron1979,Efron1986}.
Next, standard methods in maximum-likelihood estimation~\cite{Scholz2004} are employed to find the unitary matrix $W$.
Maximum-likelihood estimation simplifies under the assumption that the error on $\tilde{U}$ is a Gaussian random matrix ensemble, i.e, that the matrix entries $\left\{\tilde{U}_{ij}\right\}$ are complex independent and identically distributed (iid) Gaussian random variables centred at the correct matrix entries.
In this case, the most likely unitary matrix $W$ is the one that minimizes the Frobenius distance%
\footnote{The Frobenius norm of a matrix $A_{m \times m}$ is defined as \begin{equation}
\|A\|_F=\sqrt{\sum_{i,j=1}^m |a_{ij}|^2}.
\end{equation}
The Frobenius-norm distance between matrices $U$ and~$V$ is defined as
\begin{equation}
\mathrm{dist}(U,V) \defeq \|U-V\|_F=\sqrt{\sum_{i,j=1}^m |U_{ij}-V_{ij}|^2},
\end{equation}
and is a symmetric, positive-definite and subadditive distance function on the set of matrices.
}
from $\tilde{U}$~\cite{Tao2012}.
The unitary matrix
\begin{equation}
W = \left(\tilde{U}\tilde{U}^\dagger\right)^{-\frac{1}{2}}\tilde{U},
\label{Eq:W}
\end{equation}
minimizes the Frobenius-norm distance from $\tilde{U}$~\cite{Keller1975}.
Thus, if the random errors $\{U_{ij} - \tilde{U}_{ij}\}$ in the matrix elements are iid Gaussian random variables with mean zero, then $W$ is the best-fit unitary matrix.
Figure~\ref{Figure:UUV} is a depiction of the actual, the estimated and the most likely transformation matrices.
Algorithm~\ref{Alg:MaxLikelyUnitary} computes $W$.

This completes our procedure to estimate the most-likely unitary matrix $W$ that represents the linear optical interferometer.
In the next section, we present a procedure to estimate the error bars on the entries of the estimated representative matrix $W$ accurately.

\section{Bootstrapping to estimate error bars (Algorithm~\ref{Alg:Bootstrapping})}
\label{Sec:Bootstrapping}

In this section, we present a procedure to estimate the error bars on the matrix entries $\{W_{ij}\}$ of the characterized representative matrix $W$.
The entries $\{W_{ij}\}$ computed by Algorithms~\ref{Alg:Coincidence}--\ref{Alg:MaxLikelyUnitary} are random variables because of random error in experiments.
Obtaining accurate error bars on these random variables is important for using characterized linear optical interferometers in quantum computation and communication.
Current procedures compute error bars under the assumption that Poissonian shot noise is the only source of error in experiment~\cite{Mower2014,Carolan2015}.

We choose to employ bootstrapping on the data determine error bars~\cite{Efron1979,Efron1986,Diaconis1983,Willmott1985,Manly2006}.
Monte-Carlo simulation is widely used but this technique is not applicable here because the Poissonian shot noise assumption is not reliable given the presence of other sources of error some of which are not understood.
Bootstrapping is preferred because the nature of the error need not be characterized and instead relies on random sampling with replacement from the measured data.
Bootstrapping can be employed to yield estimators such as bias, variance and error bars.

Algorithm~\ref{Alg:Bootstrapping} calculates the error bars $\sigma(W_{ij})$ using estimates of the $\{W_{ij}\}$ pdf's, which are obtained using bootstrapping as follows.
The algorithm simulates $N$ characterization experiments using the one- and two-photon data, i.e., the inputs to Algorithms~\ref{Alg:Coincidence}--\ref{Alg:MaxLikelyUnitary}.
In each of the $N$ rounds, the one- and two-photon data are randomly sampled with replacement (resampled) to generate simulated data.
The data thus simulated are given as inputs to Algorithms~\ref{Alg:Coincidence}--\ref{Alg:MaxLikelyUnitary}, which return the simulated representative matrices 
\begin{equation}
\left\{ W'^{b}: b\in \{1,\dots,N\},\,N\in\mathds{Z}^{+} \right\}.
\end{equation}
The pdf's of the simulated-matrix entries $\{ W'^{b}_{ij}: b\in \{1,\dots,N\}\}$ converge to the pdf's of the respective elements $\{W_{ij}\}$ for large enough $N$~\cite{Diciccio1996,Davison1997}.

The simulated data are obtained in each round by resampling from the one- and two-photon experimental data as follows.
Single-photon detection counts are simulated by resampling from the set~$\{N_{ijb_{j}}:\, b_{j}\in\{1,\dots,B\}\}$ of experimental detection counts (Line~\ref{Alglin:Sim1Photon} of Algorithm~\ref{Alg:Bootstrapping}).
Two-photon coincidence counts are simulated by shuffling residuals obtained on curve-fitting experimental data.
Specifically, the algorithm (Line~\ref{Alglin:StartBootTheta}) resamples from the set 
\begin{equation}
\{r(\tau) = C^{\mathrm{exp}}_{ii'jj'}(\tau) - C_{ii'jj'}(\tau): \tau \in T\}
\end{equation}
of residuals obtained by fitting experimentally measured coincidence counts to function $C_{ii'jj'}(\tau)$~(\ref{Eq:CoincidenceRate}).
The resampled residuals are added to the fitted curve to generate the simulated data (Line~\ref{Alglin:EndBootTheta})\footnote{
The pdf of the residuals is different for different values of $\tau$.
We assume that the pdf's for different $\tau$ are of the same functional form, albeit with different widths. 
The distribution of the residuals for different values of $\tau$ are determined using standard methods for non-parametric estimation of residual distribution~\cite{Akritas2001,Chen2003}.
Algorithm~\ref{Alg:Bootstrapping} normalizes the residuals before resampling from the residual distribution.
}.
Algorithms~\ref{Alg:Coincidence}--\ref{Alg:MaxLikelyUnitary} are used to obtain the simulated elements $W_{ij}$ of the representative matrix.
Finally, the error bars on the $\{W_{ij}\}$ are estimated by the standard deviation of the pdf of the elements.

This completes the characterization of representative matrix $W$ and the error bars on its elements.
The next section details a procedure for the scattershot characterization of the interferometer to reduce the experimental time required for characterizing a given interferometer.

\begin{algorithm}[]
	\begin{algorithmic}[1]
	\Require{\Statex 
	\begin{itemize}	 
	
	\item $k, \Omega, f_1,f_2:\Omega \to \mathds{R}^+$\Comment{Spectral functions: same as Algorithm~\ref{Alg:Calibration}.}
	\item $\ell, T= \{\tau_1,\tau_2,\dots,\tau_\ell\}\in (\mathds{R}\cup 0)^\ell$\Comment{Time delay values.}
 \item $m,C^{\mathrm{cal}}(\tau), C^{\mathrm{exp}}_{ii'jj'}(\tau)$ for $\tau \in T$ and $(i,i',j,j')$ \Comment{Same as Algorithms~\ref{Alg:Calibration} and~\ref{Alg:PhaseCalcNChannel}}.
	\item $B,N_{ijb_{j}}:\{1,\dots,m\}\times\{1,\dots,m\} \times \{1,\dots,B\}\to \mathds{Z}^+$ as in Algorithm~\ref{Alg:AmplitudeEstimation}. 
 \item $N$ \Comment Number of bootstrapping samples.

	\end{itemize}
	}	
	\Ensure{\Statex
	\begin{itemize}
		\item $\sigma\left(\operatorname{re}W_{ij}\right),\sigma(\operatorname{im}W_{ij}):\{1,\dots,m\}\times\{1,\dots,m\}\to\mathds{R}^+$ \Comment{Error in $W$ elements.}
	\end{itemize}
	}
	
	\Procedure{Bootstrap}{$k, \Omega, f_1,f_2,\ell,T,\gamma,C^{\mathrm{exp}}_{ii'jj'}(\tau),\theta_{ij},B,N_{ijb},N$}
	\State $A \leftarrow \{\cos\vartheta,\sin\vartheta,\sin\vartheta,\cos\vartheta\}$, $\Phi \leftarrow \{0,\pi/2,\pi/2,0\}$
	\State $\mathrm{Residuals}^{\mathrm{cal}}(\tau) \leftarrow C^\mathrm{cal}(\tau) - \textsc{Coincidence}(k,\Omega,f_1,f_2,\ell,T,A,\Theta,\gamma)$
	\State $\mathrm{NormalResiduals}^\mathrm{cal}(\tau) \leftarrow \frac{\mathrm{Residuals}^{\mathrm{cal}}(\tau)}{C^\mathrm{fit}(\tau)}$\Comment{Assumption: $\mathrm{Residuals}^{\mathrm{cal}}(\tau)$ pdf width $\propto C^{\mathrm{fit}}(\tau)$.}
	\For{$(i,i',j,j')\in\{1,2\}\times\{1,\dots,m\}\times\{1,2\}\times\{1,\dots,m\},\,i\ne i', j\ne j'$} 
		\State $A \leftarrow \{\alpha_{i'j'},\alpha_{i'j},\alpha_{ij'},\alpha_{ij}\}$, $\Phi \leftarrow \{\theta_{i'j'},\theta_{i'j},\theta_{ij'},\theta_{ij}\}$
		\State $C^\mathrm{fit}_{ii'jj'}\leftarrow$ \textsc{Coincidence}($k,\Omega,f_1,f_2,\ell,T,A,\Theta,\gamma$)
		\State $\mathrm{Residuals}_{ii'jj'}(\tau) \leftarrow C_{ii'jj'}^\mathrm{exp}(\tau) - C_{ii'jj'}^\mathrm{fit}(\tau)$
		\State $\mathrm{NormalResiduals}_{ii'jj'}(\tau) \leftarrow \frac{\mathrm{Residuals}_{ii'jj'}(\tau) }{C_{ii'jj'}^\mathrm{fit}(\tau)}$ 
	\EndFor
	\For{$n = 1$ to $N$}
		\State $\mathrm{ShuffledNormalResiduals}^{\mathrm{cal}}(\tau) \leftarrow$ Resample $|T|$ residuals from $\mathrm{NormalResiduals}^{\mathrm{cal}}(\tau)$\label{Alglin:StartBootTheta}
		\State $\mathrm{ShuffledResidual}^{\mathrm{cal}}(\tau) \leftarrow C^\mathrm{fit}(\tau) \times \mathrm{ShuffledNormalResiduals}^{\mathrm{cal}}(\tau) $
		\State $C^{n}(\tau) = C^\mathrm{fit}(\tau) + \mathrm{ShuffledResidual}(\tau)$\label{Alglin:EndBootTheta}
		\State $\gamma^{n} \leftarrow$ \textsc{Calibration}$(k, \Omega, f_1,f_2, \ell, T,C^\mathrm{b},\vartheta)$
	\For{$(i,i',j,j')\in\{1,2\}\times\{1,\dots,m\}\times\{1,2\}\times\{1,\dots,m\},\,i\ne i', j\ne j'$} 
			\State $\alpha^{n}_{ij}\leftarrow${\sc Mean}$\sqrt{{N}_{11b_1}{N}_{ijb_j}/{N}_{1jb_j}{N}_{i1b_1}}$ from $B$ values each of $b_1,b_j$ drawn with replacement from $\{1,\dots,B\}$ \label{Alglin:Sim1Photon}
			\State $\mathrm{ShuffledNormalResiduals}_{ii'jj'}(\tau) \leftarrow |T|$ entries in $\mathrm{NormalResiduals}_{ii'jj'}(\tau)$
			\State $\mathrm{ShuffledResidual}_{ii'jj'}(\tau) \leftarrow C_{ii'jj'}^\mathrm{fit}(\tau) \times \mathrm{ShuffledNormalResiduals}_{ii'jj'}(\tau) $
			\State $C_{ii'jj'}^{n}(\tau) = C_{ii'jj'}^\mathrm{fit}(\tau) + \mathrm{ShuffledResidual}_{ii'jj'}(\tau)$	
		\EndFor
	\State $\{\theta_{ij}^{n}\}$ = {\sc ArgumentCalc}($k, \Omega,f_1,f_2,\ell, T,C^{n}_{ii'jj'}(\tau),\alpha_{ij},\gamma$)
	\State $\{W^{\prime b}_{ij}\} =\, ${\sc MaxLikelyUnitary}$\left(\left\{\alpha_{ij}^{n}\},\{\theta_{ij}^{n}\right\}\right)$
	\EndFor
	\For{$(i,j)$ in $\{1, \dots, m\}\times\{1, \dots, m\}$}
		\State $\sigma(\operatorname{re}W_{ij}) = \text{std.~dev.}\left(\{\operatorname{re}W_{ij}\}\right);\,\sigma(\operatorname{im}W_{ij}) = \text{std.~dev.}\left(\{\operatorname{im}W_{ij}\}\right)$
	\EndFor
	\EndProcedure
	
	\caption{\textsc{Bootstrap}: Estimate error bars on $\{W_{ij}\}$.}
	\label{Alg:Bootstrapping}
	\end{algorithmic}
\end{algorithm}


\section{Scattershot characterization for reduction in experimental time}
\label{Sec:Scattershot}
In this section, we present a scattershot-based characterization approach to effect a reduction in the characterization time~\cite{Lund2014,Bentivegna2015}.
Our scattershot approach reduces the time required to characterize an $m$-mode interferometer from $\BigO{m^{4}}$ to $\BigO{m^{2}}$ with constant error in the interferometer-matrix entries.

The straightforward approach of characterization involves coupling and decoupling light sources successively for each one- and two-photon measurement.
In contrast, the scattershot characterization relies on coupling heralded nondeterministic single-photon sources to each of the input ports of the interferometer and detectors to each of the output ports.
Controllable time delays are introduced at two input ports, which are labelled as the first and second ports.
All sources and detectors are switched on and the controllable time-delay values are changed first for the first port and then for the second port.

\begin{figure}[h]
\begin{centering}
\includegraphics[width=0.98\textwidth]{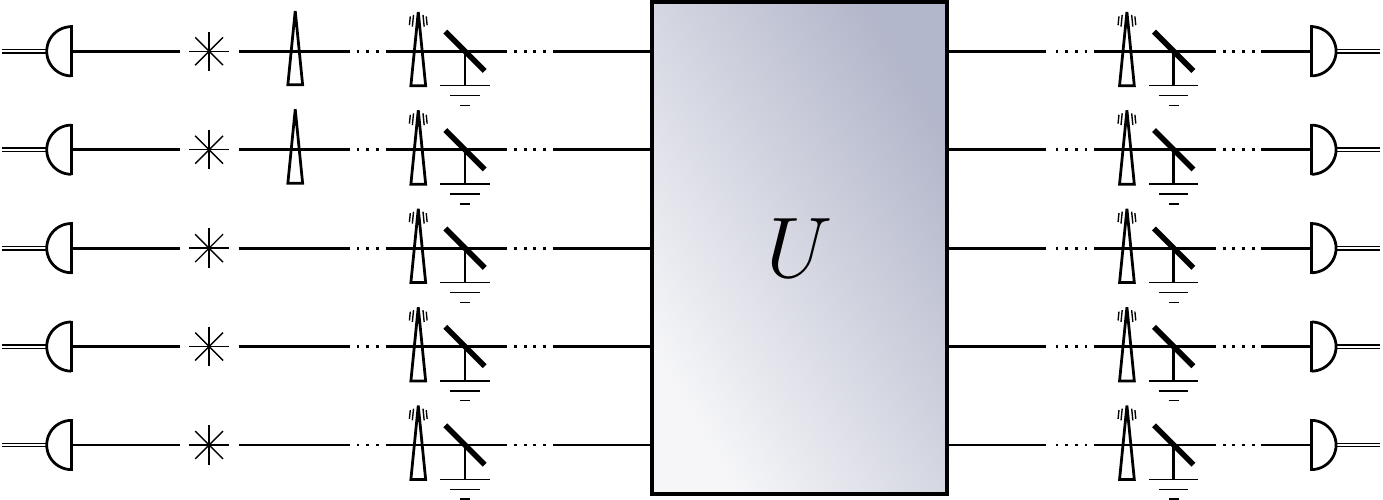}
\caption{Schematic diagram of the procedure for scattershot characterization of a five-mode interferometer $U$.
Heralded single-photon sources are coupled to the inputs of the interferometer and controllable time delays are introduced at the first two ports.
All sources and detectors are switched on and the controllable time delay values are changed for the first port and then for the second port.
The required characterization data are collected by selectively recording one- and two-photon events.
} 
\label{Figure:Scattershot}
\end{centering}
\end{figure}

Single-photon data are collected by selecting the events in which exactly one of the heralding detectors and exactly one of the output detectors register a photon simultaneously.
Two-photon coincidence events at the outputs are counted when two heralding detectors register photons.
The controllable time delays introduced at the first and second input ports ensure that each of the $2(m-1)^2$ coincidence measurements is performed.
Note that our characterization procedure (Algorithms~\ref{Alg:Coincidence}--\ref{Alg:Bootstrapping}) yields accurate estimates of interferometer parameters even when photon sources with different spectral functions are used.
In summary, the required characterization data are collected by selectively recording one- and two-photon events.
The setup for the scattershot characterization of an interferometer is depicted in Figure~\ref{Figure:Scattershot}.

Now we quantify the experimental time required in the characterization of a linear optical interferometer.
Our characterization procedure requires $Bm^2$ single-photon counting measurements and~$2(m-1)^2$ coincidence-counting measurements to characterize an $m$-mode interferometer.
We estimate the time required for each of these measurements such that random errors in the $\{\alpha_{ij}\},\{\theta_{ij}\}$ estimates remain unchanged with increasing $m$.
To ensure constant error in the $\{\alpha_{ij}\},\{\theta_{ij}\}$ estimates, we require that the number of one- and two-photon detection counts remain unchanged with increasing $m$.
The probability of photon detection at the output decreases with increasing $m$ because of the concomitant decrease in the transmission amplitudes $\{\alpha_{ij}\}$.

The amplitudes $\{\alpha_{ij}\}$ drop as $\BigO{1/\sqrt{m}}$ because of the unitarity of $U$~\cite{Dhand2014}.
Hence, one- and two-photon transmission probabilities~(\ref{Eq:PSinglePhotons},\ref{Eq:CoincidenceRate}) decrease as $1/m$ and $1/m^{2}$, respectively. 
More photons need to be incident at the interferometer input ports to offset this decrease in transmission probabilities.
Therefore, maintaining a constant standard deviation in the $\{\alpha_{ij}\}$ and~$\{\theta_{ij}\}$ measurements requires $\BigO{m}$ and~$\BigO{m^{2}}$ scaling respectively in the number of incident photons, which amounts to an overall $\BigO{m^4}$ scaling in the experimental time requirement.
Scattershot characterization allows $(m-1)^2$ different sets of the one- and two-photon data to be collected in parallel thereby reducing the time required to characterize the interferometer by a factor of $(m-1)^2$.
The overall time required for the characterization decreases from $\BigO{m^{4}}$ to $\BigO{m^{2}}$ if the scattershot approach is employed. 

Our analysis of scattershot characterization assumes that the coupling losses are small and that weak single-photon sources are used, i.e., that the probability of multi-photon emissions from the heralded sources is small as compared to single-photon emission probabilities.
These assumptions are expected to hold for on-chip implementations of linear optics that have integrated single-photon sources and detectors.

Light sources used at each input port in our scattershot-based characterization procedure differ spectrally in generally.
Our characterization procedure is accurate despite this difference because we measure source-field spectra and using these data in the curve-fitting procedure.

We have developed the scattershot approach which has advantages and disadvantages but on balance is a superior experimental approach to consecutive measurement.
The advantage is that the time requirement for characterization if reduced by a factor that scales as~$\BigO{m^{2}}$.
The disadvantage is the overhead of requiring one source at each input port and one detector at each output port.
The disadvantage is not daunting because these requirements are commensurate with other active investigations of QIP such as LOQC and scattershot BosonSampling.
In fact, state of the art implementations~\cite{Bentivegna2015} meet our increased requirements for scattershot characterization.

\section{Summary of procedure and discussions}
\label{Sec:Discussions}
In this section, we summarize our characterization procedure for a less formally-oriented audience.
We describe the processing of the collected experimental data by the various algorithms presented in Section~\ref{Sec:Procedure}.
We compare our procedure with the existing procedure for the characterization of linear optics using one- and two-photons~\cite{Laing2012}.
We provide numerical evidence that our characterization procedure promises enhanced accuracy and precision even in the presence of shot noise and mode mismatch.

The experimental data required by our procedure to characterize an $m$-mode interferometer includes the following one- and two-photon measurements.
The number $N_{ijb_{j}}$~(\ref{Eq:Nijbdef}) of single-photon detection events is counted at the $j$-th output port when single photons are incident at the $i$-th input port.
This single-photon counting is repeated $B$ times for each of the input ports and output ports, where $B$ is chosen such that the cumulants of the set $\{N_{ijb_{j}}:b_{j}\in\{1,\dots,B\}\}$ converge.
The single-photon counts $\{N_{ijb_{j}}\}$ are received by Algorithm~\ref{Alg:AmplitudeEstimation}, which returns the $\{\alpha_{ij}\}$~(\ref{Eq:RMS0}) estimates using Eq.~(\ref{Eq:AlphaMeasure}).

The spectral function $f_{j}(\omega)$~(\ref{Eq:Single}) of the light incident at each input port $j$ is measured.
This function is used by Algorithm~\ref{Alg:Coincidence} to calculate the expected two-photon coincidence curves using Eq.~\ref{Eq:CoincidenceRate}.
Fitting experimental data to these coincidence curves yields an accurate estimate of the mode-matching parameter during calibration and the arguments $\{\theta_{ij}\}$ in the argument-estimation procedure.
Thus, the spectral function $f_{j}(\omega)$ serves as an input to the algorithms for the estimation of the mode-matching parameter
 and of the arguments $\{\theta_{ij}\}$ (Algorithms~\ref{Alg:Calibration}--\ref{Alg:PhaseCalcNChannel}).

The mode-matching parameter $\gamma$ is estimated by performing coincidence measurement on a beam splitter that is separate from the interferometer but is constructed using the same material.
First, we use single-photon data to estimate the reflectivity $\cos\vartheta$ of the beam splitter according to Eq.~(\ref{Eq:FindVartheta}).
Imperfect mode-matching changes the shape of the coincidence curve, and we find $\gamma$ by comparing the shapes of (i) the curve expected for reflectivity $\cos\vartheta$ and (ii) the curve obtained experimentally.
The estimated beam splitter reflectivity, the measured spectra and the coincidence counts are received as inputs by Algorithm~\ref{Alg:Calibration}, which returns an estimate of $\gamma$.

Algorithm~\ref{Alg:PhaseCalcNChannel} uses two-photon coincidence counts to estimate the arguments $\{\theta_{ij}\}$.
Coincidence counts are measured for the input ports $j,j'$ and output ports $i,i'$ for the $2(m-1)^{2}$ sets 
\begin{equation}
(i,i',j,j') \in \{1,2\}\times\{1,\dots,m\}\times \{1,2\}\times\{1,\dots,m\}, \quad i\ne i',\, j\ne j'
\label{Eq:Choice}
\end{equation}
 of input and output ports.
In other words, coincidence counts are measured for different choices of two input ports and two output ports, such that each of the choices includes (i) either the first or the second input ports and (ii) either the first or the second output port. 
Algorithm~\ref{Alg:PhaseCalcNChannel} receives as input the measured spectra, the $\{\alpha_{ij}\}$ values estimated by Algorithm~\ref{Alg:AmplitudeEstimation}, the $\gamma$ value estimated by Algorithm~\ref{Alg:Calibration} and the two-photon coincidence data for the choice~(\ref{Eq:Choice}) of input ports.
The algorithm returns the $\{\theta_{ij}\}$ estimates.
The computed estimates of $\{\alpha_{ij}\}$ and of $\{\theta_{ij}\}$ yield the representative unitary matrix $W$ (\ref{Eq:W}) that has maximum likelihood of describing the characterized interferometer (Algorithm~\ref{Alg:MaxLikelyUnitary}).
This completes a summary of our procedure for characterization of the interferometer.

Algorithm~\ref{Alg:Bootstrapping} employs bootstrapping to find the error bars on the elements $\{W_{ij}\}$ of the characterized unitary matrix. 
The bootstrapping procedure uses the experimental data that is received by Algorithms~\ref{Alg:Coincidence}--\ref{Alg:MaxLikelyUnitary} and repeatedly simulates experiments by resampling from the experimental data. 
The number $N$ of repetitions is chosen such that the pdf's of the $\{W_{ij}\}$ elements over many rounds of simulation converge.
The error bars on the $\{W_{ij}\}$ elements are computed based on the estimated pdf's of the elements. Our procedure thus enables the estimation of meaningful error bars on the characterized unitary matrix.

Bootstrapping is employed to test the goodness of fit between the experimental curve and expected curves~\cite{Kojadinovic2012}.
Experiments~\cite{Metcalf2013,Tillmann2014} can employ bootstrapping instead of the incorrect $\chi^{2}$-confidence measure to test if the data are consistent with quantum predictions or with the classical theory.

Finally, we recommend a scattershot approach for reducing the experimental time required to characterize interferometers.
The approach involves coupling heralded nondeterministic single-photon sources at each of the input ports and single-photon detectors at each of the output ports.
All the sources and the detectors are switched on in parallel.
Single-photon counts are recorded selectively as two-photon coincidences between the heralding detectors and the output detectors, while two-photon events are recorded when two heralding detectors and two output detectors record photons.
Controllable time delays are introduced at the first and second input ports so coincidences between each of the $2(m-1)^2$ choices~(\ref{Eq:Choice}) of input and output ports are recorded.
The scattershot approach reduces the experimental time required to characterize an $m$-mode interferometer from $\BigO{m^{4}}$ to $\BigO{m^{2}}$.

Now we compare and contrast our procedure with the Laing-O'Brien procedure~\cite{Laing2012}.
Our procedure is inspired by the Laing-O'Brien procedure in that it employs
(i) a `real-bordered' parameterization~(\ref{Eq:RMS0}) of the representative matrix and modelling of linear losses at the interferometer ports, 
(ii) a ratio of single-photon data to estimate the complex amplitudes of the matrix elements and
(iii) an iterative procedure that uses two-photon data to estimate the amplitudes of the complex arguments and to estimate the signs of the complex arguments.

Our procedure differs from the Laing-O'Brien~\cite{Laing2012} procedure in that we use averaged value~(\ref{Eq:AlphaMeasure}) of the ratio of single-photon detections over many runs rather than the ratio of averaged values. 
This difference ensures accuracy of our procedure even under fluctuation in the number of incoming photons. 
Such fluctuations might arise from fluctuations in pump strength of the single photon source or in the strength of coupling between photon source and interferometer.

\begin{figure}[h]
\begin{centering}
\includegraphics[width=0.7\textwidth]{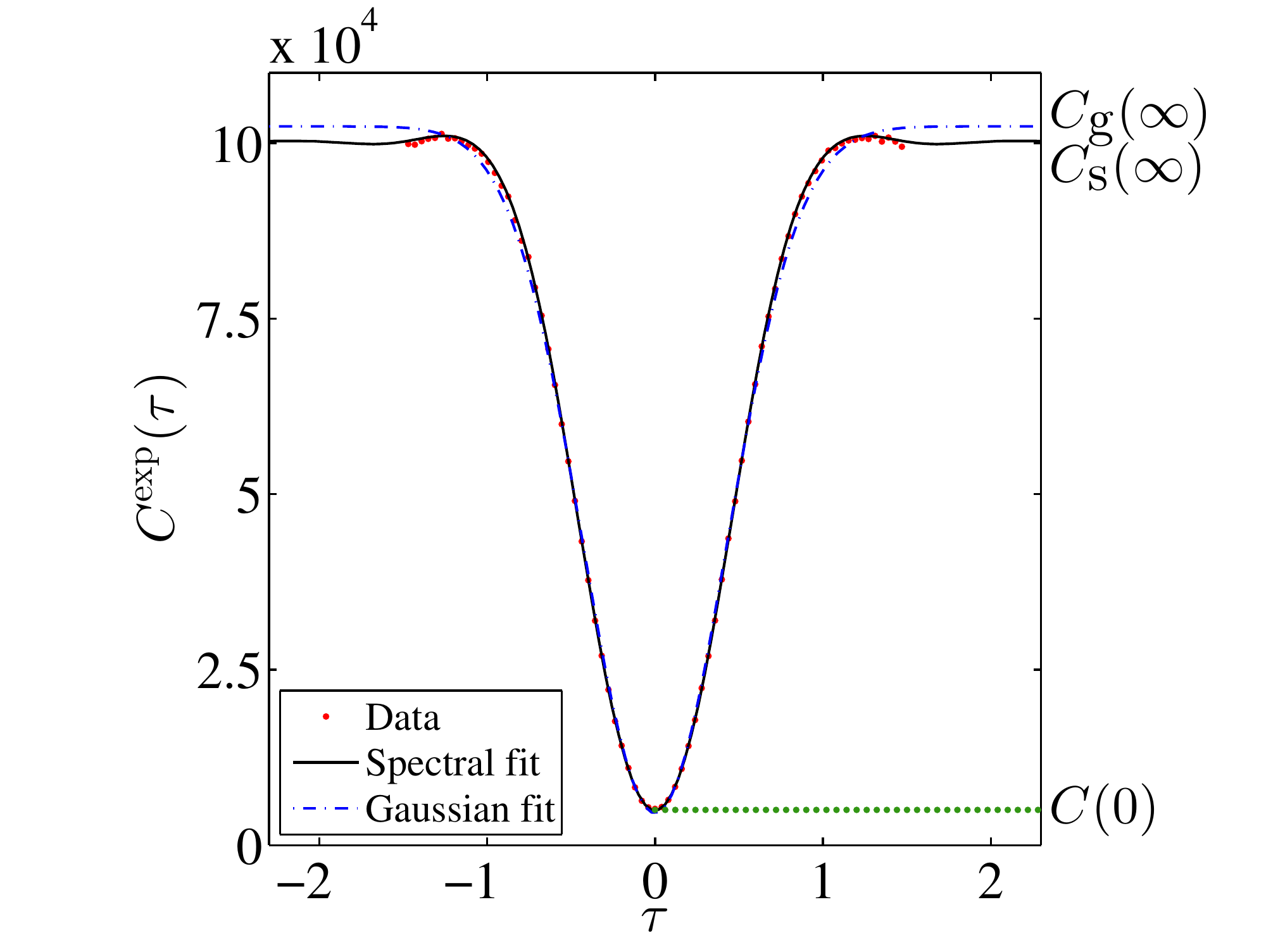}
\caption{
The fitting of coincidence data to curves obtained from spectral functions using~(\ref{Eq:CoincidenceRate}) and to Gaussian functions. 
Coincidence counts are simulated using experimentally measured spectra.}
\label{Figure:CoincidenceCurveFit}
\end{centering}
\end{figure}

\begin{figure}[h]
\begin{center}
\includegraphics[width=0.7\textwidth]{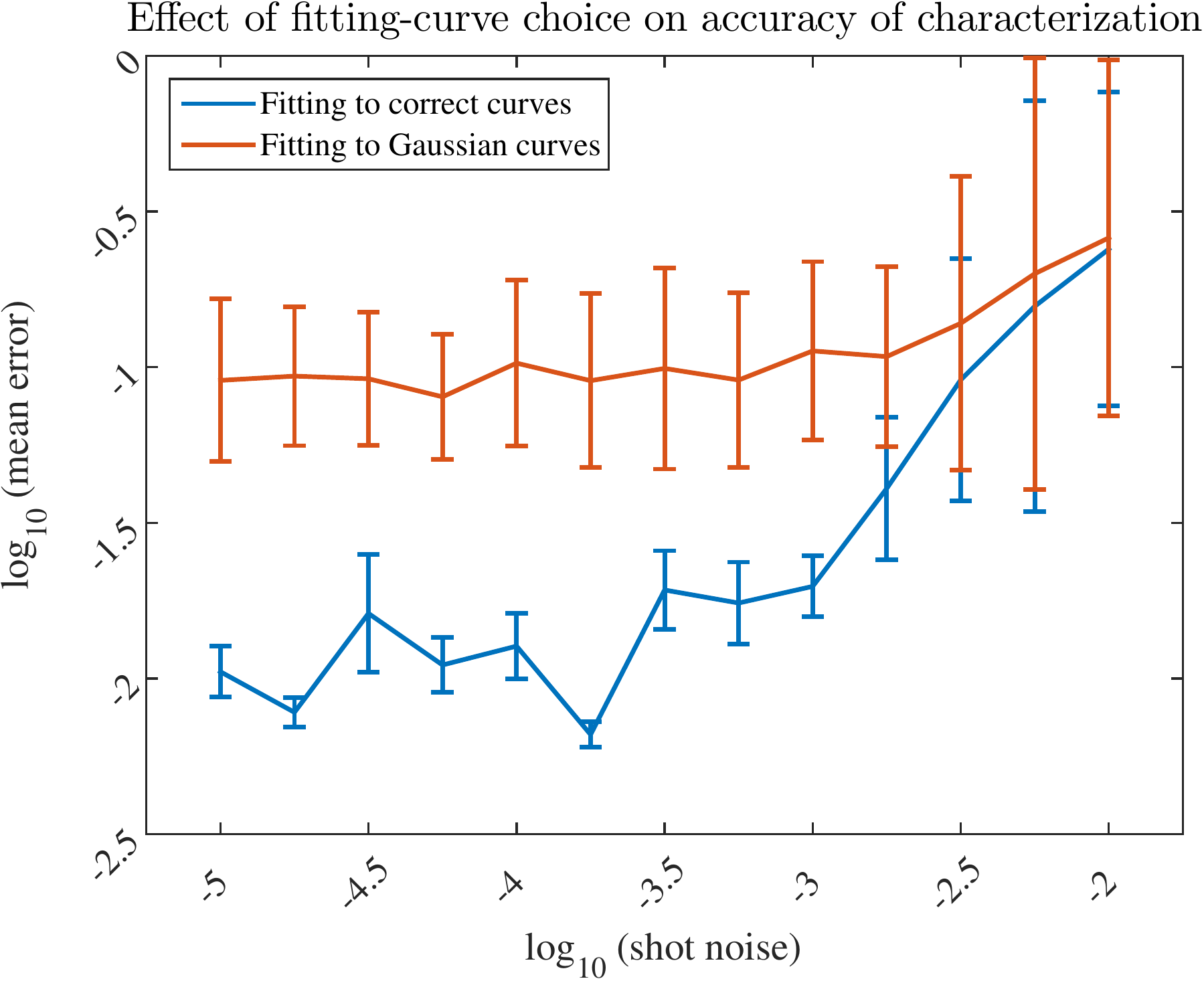}
\caption{A plot showing the effect of fitting-curve choice on the accuracy and precision of the characterization procedure. The two curves depict the mean error for the two different choices of fitting curves, where the error is the trace distance between the expected and the actual unitary transformations and the mean is over 1000 simulated characterization experiments. 
One- and two-photon interference data was simulated for a five-channel interferometer using experimentally measured spectra and simulated Poissonian shot-noise.
Characterization was performed by fitting coincidence curves to Gaussians (red curve) and to correct curves according to our procedure (blue curve).
\textsc{matlab} code for the simulations depicted in this figure is available on GitHub~\cite{Dhand2015}
}
\label{Figure:GaussianVersusNongauss}
\end{center}
\end{figure}

Another advance in our method is the curve-fitting procedure for estimating complex arguments of interferometer matrices. 
The Laing-O'Brien procedure requires coincidence-curve visibilities to estimate complex arguments~$\alpha_{ij}$.
Whereas the Laing-O'Brien procedure recommends coincidence probabilities be measured at zero time delay and also at time delays large as compared to the temporal spread of the wave-packet, in practice, current implementations determine the visibilities by fitting experimental data to Gaussian curves~\cite{Hong1987,Peruzzo2011,Crespi2013,Spring2013,Tillmann2013,Carolan2014}.
These implementations are flawed because source spectra differ from Gaussian in general.
Our procedure is accurate because the data are fit to curves computed from spectral functions, rather than fitting to Gaussians. 
Figure~\ref{Figure:CoincidenceCurveFit} illustrates the distinction between fitting experimental coincidence counts to the coincidence function~(\ref{Eq:CoincidenceRate}) simulated using spectra and fitting to Gaussian functions.
Figure~\ref{Figure:GaussianVersusNongauss} demonstrates the increase in accuracy and precision of characterization by using the correct curve-fitting function.

\begin{figure}[h]
\begin{center}
\includegraphics[width=0.6\textwidth]{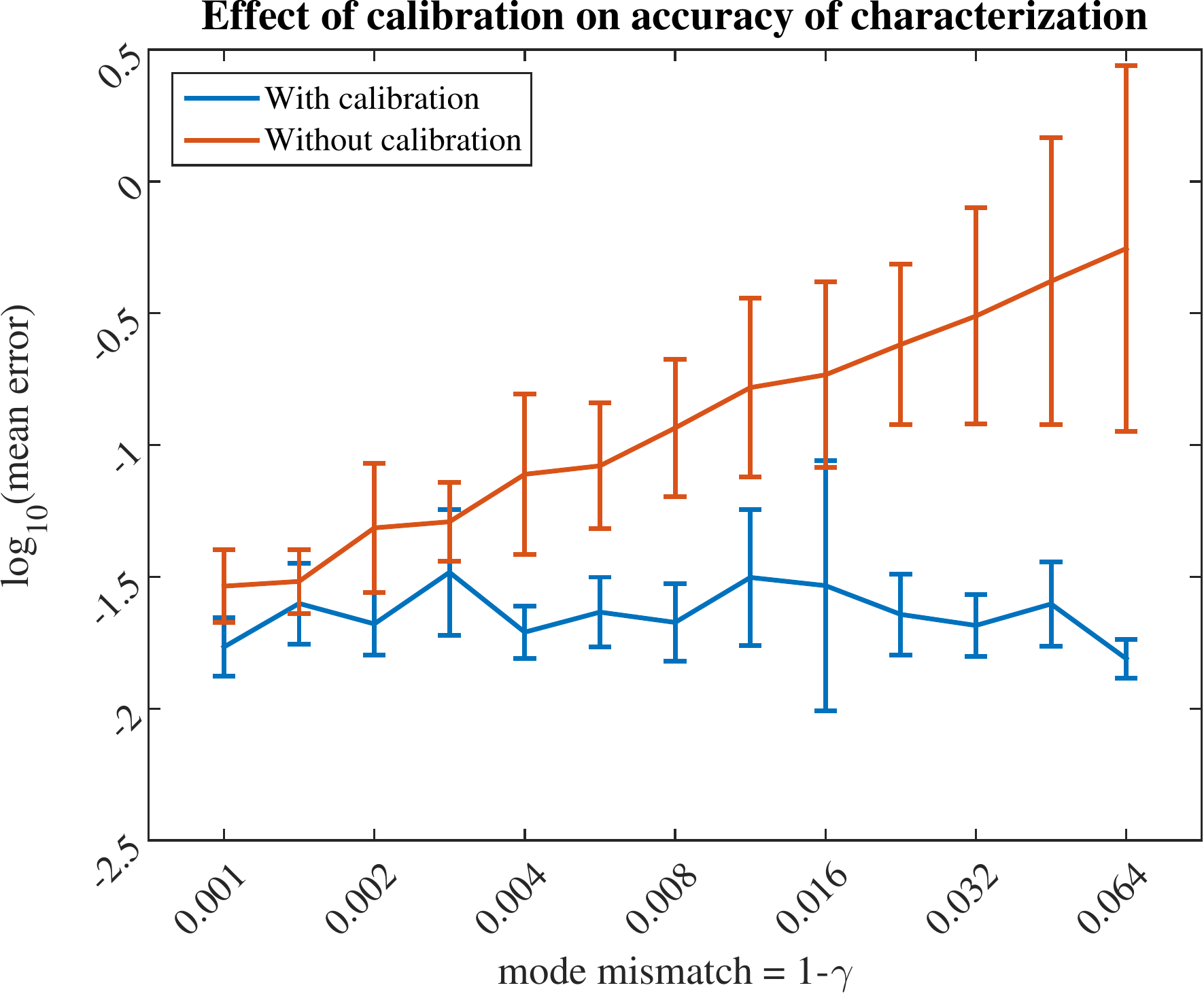}
\caption{A plot showing the effect of calibration on the accuracy and precision of the characterization procedure. The two curves depict the mean error for characterization with (blue curve) and without calibration (red curve), where the error is the trace distance between the expected and the actual unitary transformations and the mean is over 1000 simulated characterization experiment. The simulations comprised generating one- and two-photon interference data based on experimentally measured spectral functions and performing characterization by our procedure.
\textsc{matlab} code for the simulations depicted in this figure is available on GitHub~\cite{Dhand2015}
}
\label{Figure:IsCalibrationUseful}
\end{center}
\end{figure}

We introduce the calibration subroutine, which relies on the estimation of the mode mismatch in the source field.
Spatial and polarization mode mismatch is not an issue of major concern in waveguide-based interferometers, which typically operate in the single-photon regime.
In these interferometers, the calibration step of our procedure can be neglected without decreasing accuracy.
The mode-mismatch parameter~$\gamma$, which is an input of the curve-fitting procedure, is set to unity.

In the context of bulk-optics, our calibration step ensures accuracy and precision if (i)~$\gamma$ is identified as the maximum-possible source overlap in the spatial and polarization degrees of freedom and (ii) the experimentalist adjusts the setup to maximize coincidence visibility for the calibrating beam splitter and for each choice of interferometer inputs ports.
Such an adjustment will ensure that the source overlap acquires its maximum-possible value $\gamma$ in each of the coincidence-curve measurements.
This maximum value is a property of the sources used and is independent of source alignment and focus so is expected to remain unchanged between different confidence measurements.
Figure~10 demonstrates the increase in accuracy and precision of characterization by using the calibration procedure

Other advances made in our characterization procedure over existing procedures include (i) a maximum likelihood estimation approach to determine the unitary matrix that best fits the data (ii) a bootstrapping based procedure to obtain meaningful estimates of precision and (iii) a scattershot-based procedure to improve the experimental requirements of characterization.

\section{Conclusion}
\label{Sec:Conclusion}
In conclusion, we devise a one- and two-photon interference procedure to characterize any linear optical interferometer accurately and precisely.
Our procedure provides an algorithmic method for recording experimental data and computing the representative transformation matrix with known error. 

The procedure accounts for systematic errors due to spatiotemporal mode mismatch in the source field by means of a calibration step and corrects these errors using an estimate of the mode-matching parameter.
We measure the spectral function of the incoming light to achieve good fitting between the expected and measured coincidence counts, thereby achieving high precision in characterized matrix elements.
We introduce a scattershot approach to effect a reduction in the experimental requirement for the characterization of interferometer.
The error bars on the characterized parameters are estimated using bootstrapping statistics.

Bootstrapping computes accurate error bars even when the form of experimental error is unknown and is, thus, advantageous over the Monte Carlo method.
Hence, our bootstrapping-based procedure for estimating error bars can replace the Monte Carlo method used in existing linear-optics characterization procedures.
We open the possibility of applying bootstrapping statistics for the accurate estimation of error bars in photonic state and process tomography.

\section*{Acknowledgments}
We thank Matthew A.~Broome, Jens Eisert, Dirk R.~Englund, Sandeep K.~Goyal, Hubert de Guise and Michael J.~Steel for useful comments.
ID and BCS acknowledge AITF, China Thousand Talent Plan and NSERC and USARO for financial support.
AK acknowledges CryptoWorks21 and NSERC for financial support.

\pagebreak

\appendix
\section{Removal of instability in characterization procedure}
\label{Sec:Instability}
In this section, we describe an instability in our characterization procedure, which can yield large error in the $\{W_{ij}\}$ output for small error in the experimental data $C^{\mathrm{exp}}_{ii'jj'}(\tau)$ in case of certain interferometers $W$.
We present a strategy to circumvent this instability by means of collecting and processing additional experimental data.

The instability in the characterization procedure arises because of an instability in estimation of $\{\operatorname{sgn}\theta_{ij}\}$ (Algorithm~\ref{Alg:SignCalc}).
Small error in the measured coincidence counts can lead to the wrong inference of $\operatorname{sgn}\theta_{ij}$, which can lead to a large error $\|W-U\|$ in the characterized matrix $W$.
Recall that Algorithm~\ref{Alg:SignCalc} uses the identity $\operatorname{sgn}\theta_{ij} = \operatorname{sgn}\left(\big|\beta_{ii'jj'}-\beta_{ii'jj'}^{+}\big|-\big|\beta_{ii'jj'}-\beta_{ii'jj'}^{-}\big|\right)$ (\ref{Eq:SignEquation}) to determine the sign of the arguments, where $\beta^{\pm}_{ii'jj'}\defeq |\theta_{i'j'}-\theta_{ij'}-\theta_{i'j}\pm|\theta_{ij}||$ and the values of $\beta_{ii'jj'}, \theta_{i'j'},\theta_{i'j},\theta_{ij'},\left|\theta_{ij}\right|$ are estimated by curve fitting.

Random and systematic error in measured coincidence counts can lead to estimate of variables $\beta_{ii'jj'}, \theta_{i'j'},\theta_{i'j},\theta_{ij'},\left|\theta_{ij}\right|$ differing from their actual values.
The estimation of $\operatorname{sgn}\theta_{ij}$ is unstable if the $\theta_{i'j'}-\theta_{i'j}-\theta_{ij'}$ term (\ref{Eq:SignEquation}) is close to $0$ or $\pi$ because, in this case, a small error in the $\beta_{ii'jj'}$ estimate can lead to an incorrect $\operatorname{sgn}\theta_{ij}$ estimate.
In other words, the sign estimates are unstable if the values of
\begin{equation}
\theta_{ii'jj'}^{\mathrm{ref}} = \mathrm{min}\left[\theta_{i'j'} - \theta_{ij'} - \theta_{i'j},\pi-(\theta_{i'j'} - \theta_{ij'} - \theta_{i'j})\right],
\end{equation}
are small compared to the error in our $\beta_{ii'jj'}, \theta_{i'j'},\theta_{i'j},\theta_{ij'},\left|\theta_{ij}\right|$ estimates.

\begin{figure}[H]
\begin{centering}
\hspace{2ex}
 \begin{subfigure}{0.51\textwidth}
 \includegraphics[width=\textwidth]{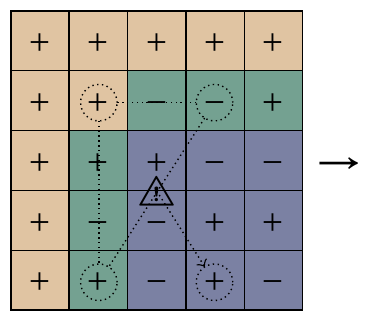}
 \caption{}
 \end{subfigure}
 \begin{subfigure}{0.43\textwidth}
 \includegraphics[width=\textwidth]{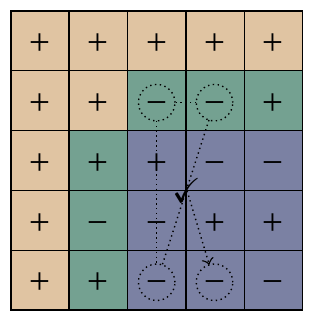}
 \caption{} 
 \end{subfigure}
\end{centering}
\caption{An illustration of the instability in the $\theta_{ij}$ characterization procedure for an interferometer with $m=5$ modes.
(a)~If the value of $|\theta_{22}-\theta_{52}-\theta_{24}|$ is close to $0$ or $\pi$, then small error in $C_{2524}^{\mathrm{exp}}(\tau)$ can lead to an error in the estimation of $\operatorname{sgn}(\theta_{54})$.
(b)~The instability in the $\theta_{54}$ can be removed by collecting two-photon coincidence data for output ports $2,5$ and input ports $3,4$ and using the values of $\theta_{23},\theta_{53},\theta_{24}$ instead of $\theta_{22},\theta_{52},\theta_{24}$ values.}
\label{Figure:Syndrome}
\end{figure}

We mitigate the sign-inference instability by making two modifications to our characterization procedure; the first modification removes instability from the sign-inference of the second row and second column elements whereas the second modification prevents incorrect inference of the remaining signs. 
The inference of $\{\operatorname{sgn}\theta_{i2}\},\{\operatorname{sgn}\theta_{2j}\}$ (Lines~\ref{Alglin:StartSecondRowColumn}--\ref{Alglin:EndSecondRowColumn}, Figures~\ref{Figure:Ordering}b,~\ref{Figure:Ordering}c) is unstable if 
\begin{equation}
\theta_{i2j2}^{\mathrm{ref}} =\mathrm{min}(\theta_{22},\pi-\theta_{22})
\end{equation}
is small as compared to the error in the $\beta_{i2j2}, \theta_{22},\theta_{2j},\theta_{i2},\left|\theta_{ij}\right|$ estimates.
Hence, we relabel the interferometer ports such that $\theta_{22}$ is as far away from $0$ and $\pi$ as possible.
Specifically, after the amplitudes of the phases have been estimated (Line~\ref{Alglin:EndAmplitudes} of Algorithm~\ref{Alg:PhaseCalcNChannel}), we choose $i,j$ for which $|\theta_{ij}-\pi/2|$ is minimum, and we swap the labels of input ports $2,j$ and output ports $2,i$.
We measure two-photon coincidence counts based on this new labelling and process it using Algorithm~\ref{Alg:PhaseCalcNChannel}.
The instability in the procedure for estimation of the $\{\theta_{i2}\},\{\theta_{2j}\}$ signs is removed as a result of the relabelling.

The second modification is aimed at removing the instability in the remaining signs.
The procedure estimates the remaining signs by using $\{C^{\mathrm{exp}}_{ii'jj'}(\tau)\}$ values for $i'=j'=2$.
The estimation of $\theta_{ij}$ is unstable if $\theta_{i2j2}^{\mathrm{ref}}$ is small as compared to the error in the $\beta_{i2j2}, \theta_{22},\theta_{2j},\theta_{i2},\left|\theta_{ij}\right|$ estimates.
 We make a heuristic choice of a threshold angle $\theta^{\mathrm{T}}$ that accounts for the error in these variables, and we reject any $\operatorname{sgn}\theta_{ij}$ inferred using $ \theta_{i2j2}^{\mathrm{ref}} \le \theta^{\mathrm{T}}$.
 Additional two-photon coincidence counting is performed and employed to estimate these values of $\theta_{ij}$, as detailed in the following lines that can be added to the algorithm to remove the instability
\begin{algorithmic}[1]
 \setcounter{ALG@line}{0}
 \algrenewcommand{\alglinenumber}[1]{\footnotesize{\ref{Alglin:EndSecondRowColumn} + #1:}}
 \For{$(i,j)$ in $\{3, \dots, m\}\times\{3, \dots, m\}$}
 \algrenewcommand{\alglinenumber}[1]{\footnotesize{\hspace{2em} #1:}}
 \If{$\theta^r_{i2j2} < \theta_T$}
 \State Choose $i'\ne 1,i$ and $j'\ne 1,j$ such that $|\theta_{i'j'}-\theta_{ij'}-\theta_{j'j}|$ is closest to $\pi/2$.	
 	\State $C^{\mathrm{exp}}_{ii'jj'}(\tau)\leftarrow$ Coincidence counts for input ports $j,j'$ and output ports $i,i'$. 
 \State $A \leftarrow \{0,0,0\}, \Phi \leftarrow \{0,0,0\}$
		\State $\beta_{ii'jj'} \leftarrow \textsc{Argument2Port}(C^{\mathrm{exp}}_{ii'jj'}(\tau),\Omega, f_1, f_2,T,A,\Phi,\gamma)$
		\State $\theta_{ij} \leftarrow \theta_{ij}${\sc SignCalc}$(\beta_{ii'jj'},\theta_{i'j'},\theta_{ij'},\theta_{i'j},|\theta_{ij}|)$ 
 \EndIf
	\EndFor
\end{algorithmic}
Figure~\ref{Figure:Syndrome} illustrates the rejection of those $i,j$ choices for which $\theta_{i2j2}^{\mathrm{ref}} \le \theta^{\mathrm{T}}$ and the use of $ C^{\mathrm{exp}}_{ii'jj'}(\tau), j'\ne 2$ counts to obtain a correct estimate of $\operatorname{sgn}\theta_{ij}.$
We thus remove the instability in the estimation of $\{\theta_{ij}\}$ and in the estimation of the representative matrix $W$.

\section{Curve-fitting subroutine}
\label{Sec:CurveFitting}
Our characterization procedure employs curve fitting in Algorithm~\ref{Alg:Calibration} to estimate the mode-matching parameter~$\gamma$ and in Algorithms~\ref{Alg:PhaseCalc2Channel}--\ref{Alg:PhaseCalcNChannel} to estimate $\{\theta_{ij}\}$ values.
The curve-fitting procedure determines those values of unknown parameters that maximize the fitting between experimental and expected coincidence data.
In this section, we describe the inputs and outputs of the curve-fitting subroutines.
We present heuristics to compute good initial guesses of the fitted parameters. 

\begin{figure}
\centering
\hspace{-5ex}
\begin{subfigure}{0.52\textwidth}
 \includegraphics[width=1.03\textwidth]{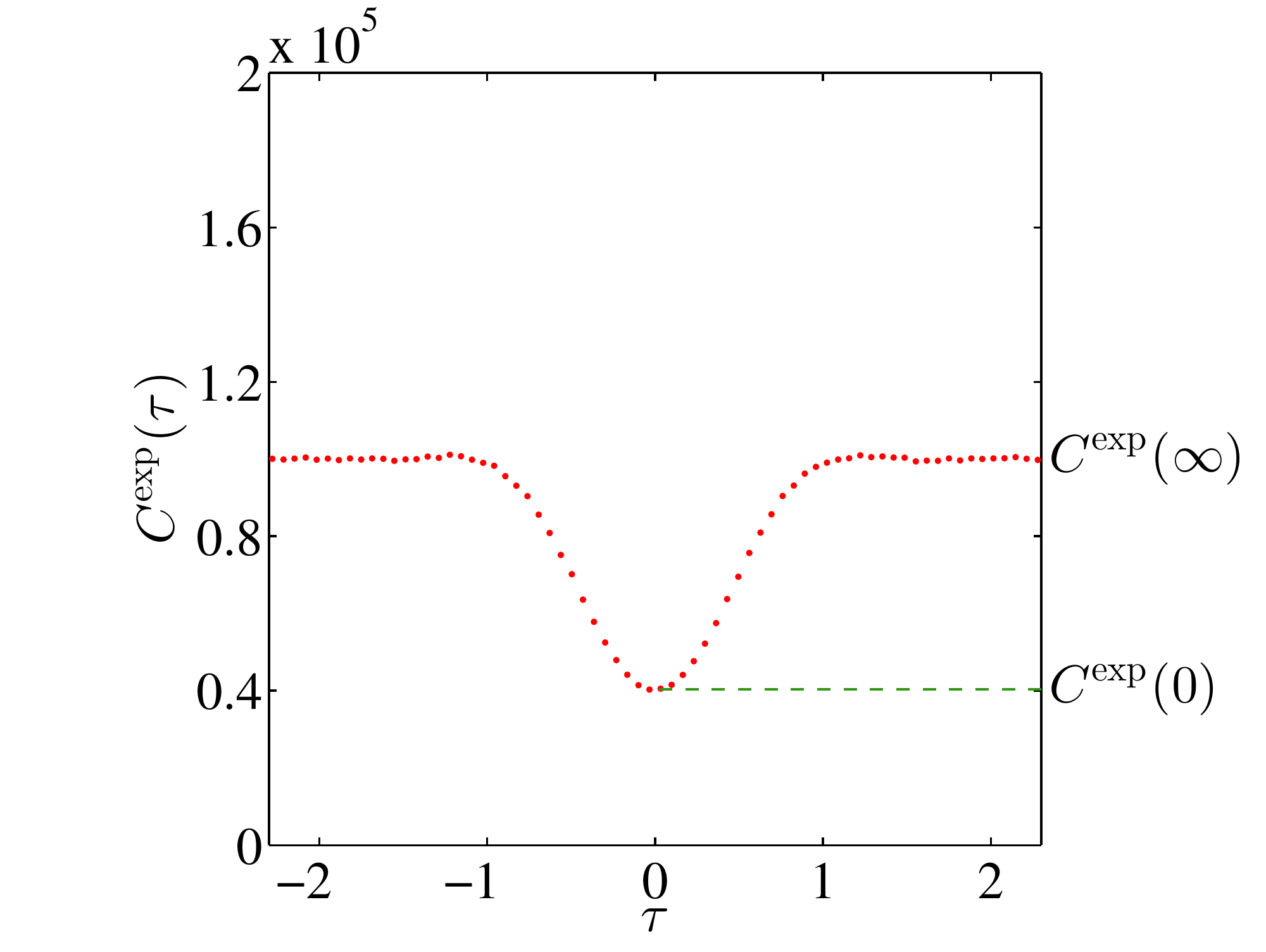}
 \caption{}
 \label{Figure:CoincidenceShapeSU2}
\end{subfigure}
\begin{subfigure}{0.52\textwidth}
 \includegraphics[width=1.03\textwidth]{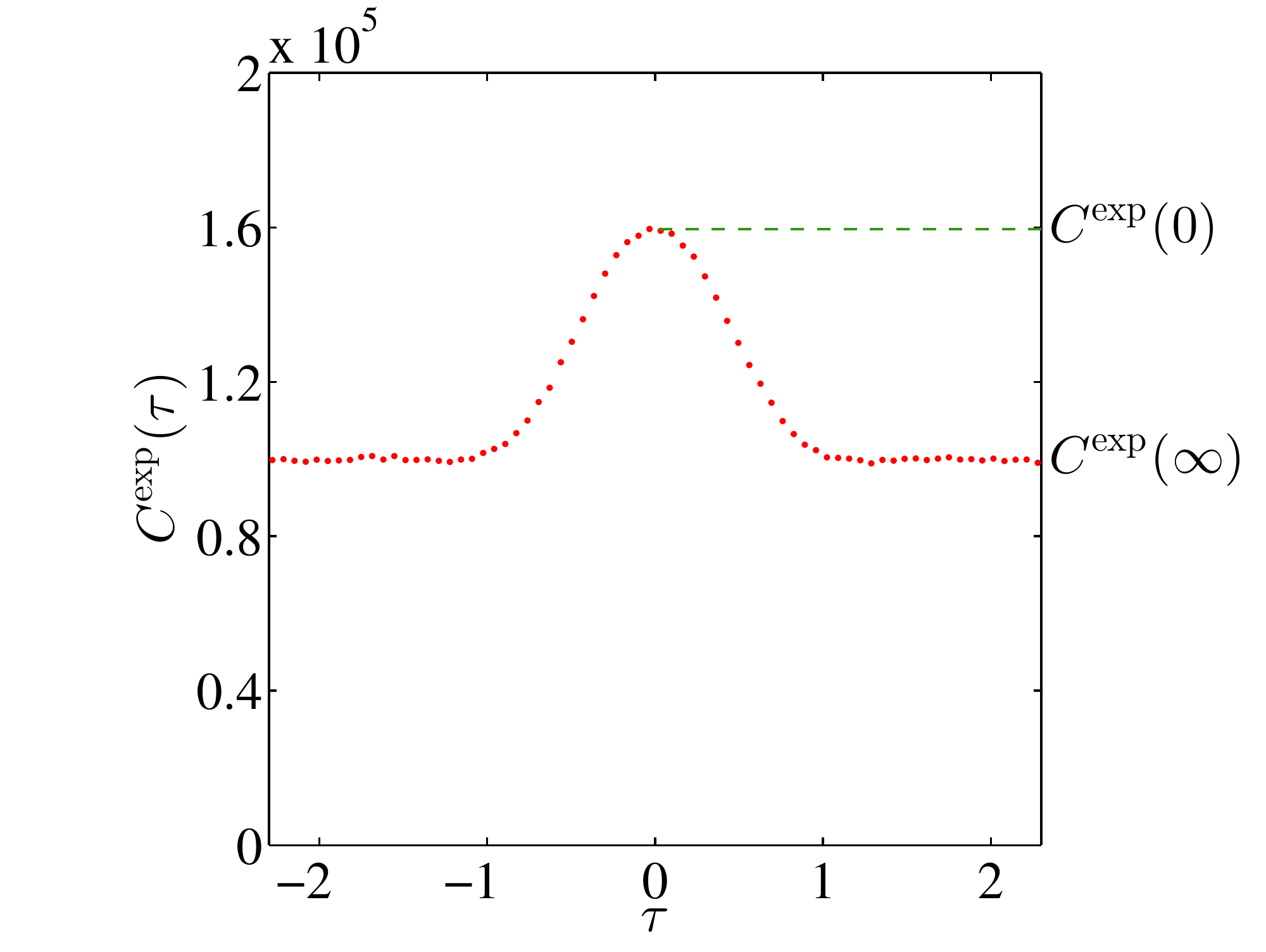}
 \caption{}
 \label{Figure:CoincidenceShapeBetaZero}
\end{subfigure}

\hspace{-5ex}
\begin{subfigure}{0.52\textwidth}
 \includegraphics[width=1.03\textwidth]{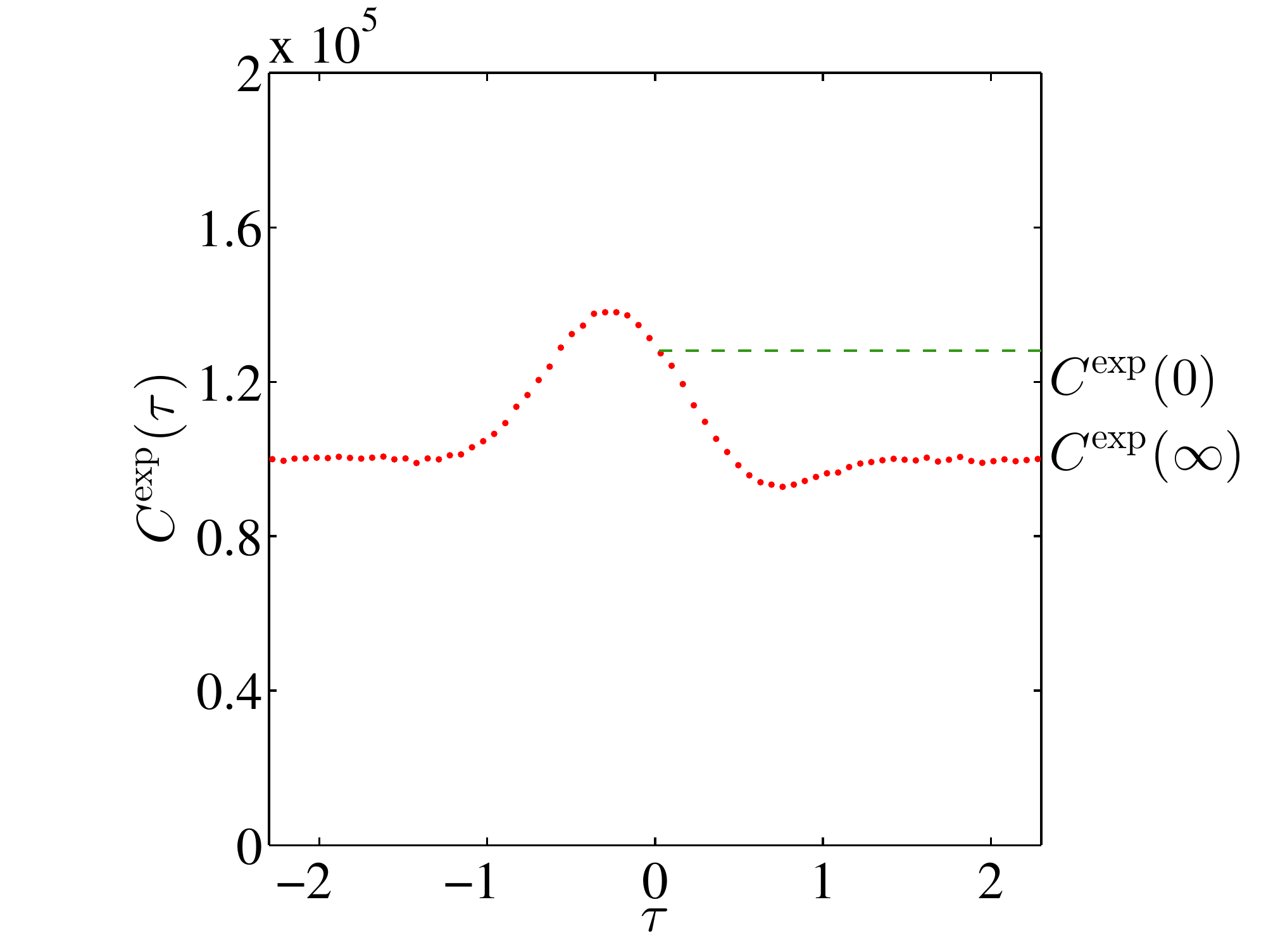}
 \caption{}
 \label{Figure:CoincidenceShapeGeneral1}
\end{subfigure}
\begin{subfigure}{0.52\textwidth}
 \includegraphics[width=1.03\textwidth]{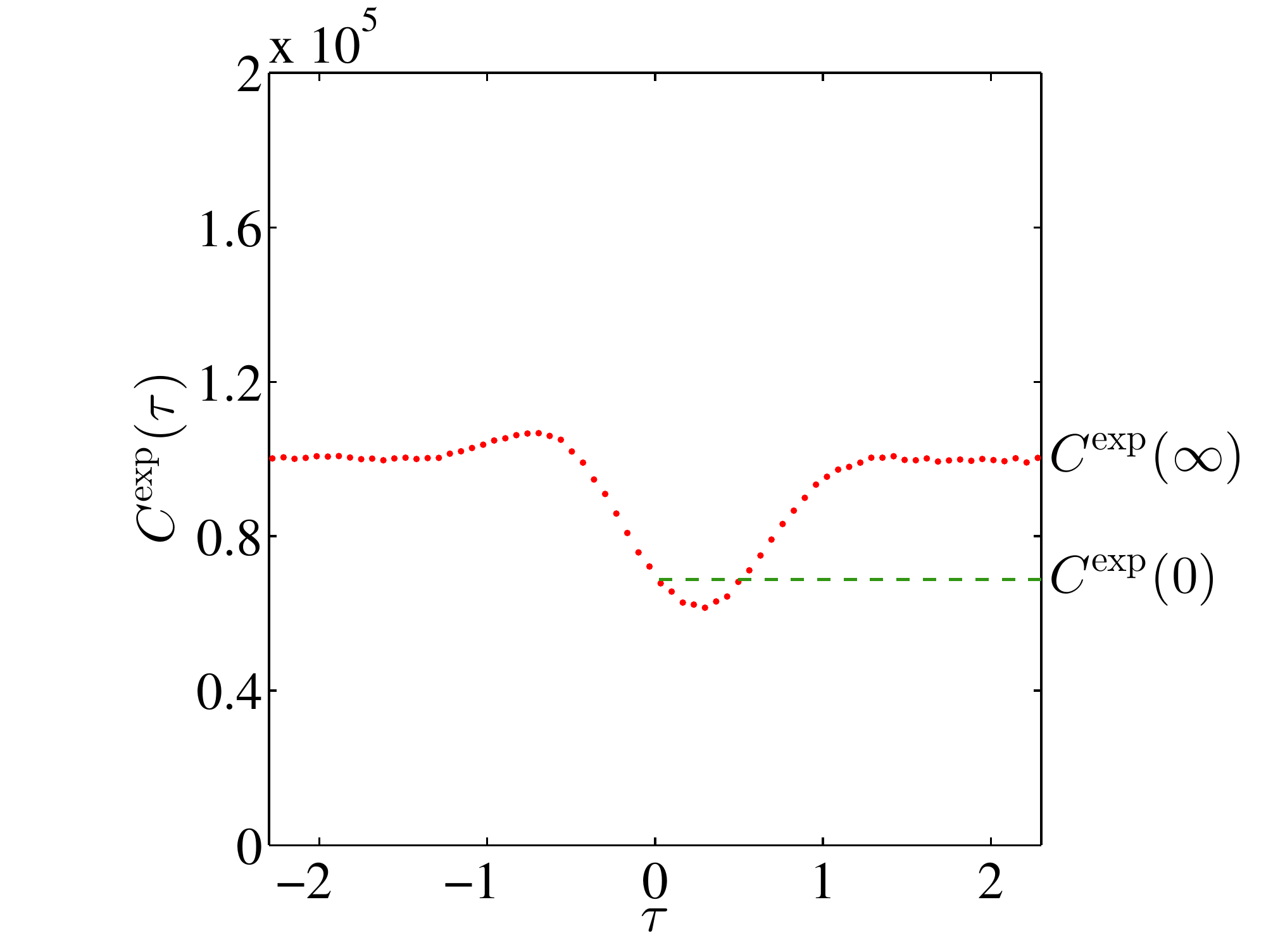}
 \caption{}
 \label{Figure:CoincidenceShapeGeneral2}
\end{subfigure}
\caption{Simulated coincidence counts for output ports $i,i'$ and input ports $j,j'$ of interferometer with $\alpha_{ii}=\alpha_{i'j'}=\sqrt{3}/4$ and $\alpha_{ii'}=\alpha_{ij'}=1/4$ and for different values of $\beta_{ii'jj'}$. The value of $\beta_{ii'jj'}$ in each respective figure is (a) $\pi$, (b) $0$, (c) ${\pi}/{3}$ and (d) ${2\pi}/{3}$. The coincidence counts corresponding to $\tau=0$ and $\tau\to\infty$ are marked on each plot by $C^{\mathrm{exp}}(0)$ and $C^\mathrm{\text{exp}}(\infty)$ respectively.
}
\label{Figure:CoincidenceCurveShapes}
\end{figure}

The curve-fitting subroutine receives as input 
(i) the choice of parameters to be fitted; 
(ii) the coincidence counts $\{C_{ii'jj'}^{\text{exp}}(\tau)\}$; 
(iii) an objective function, which characterizes the least-square error between expected and experimental counts; and 
(iv) the initial guesses for each of the fitted parameters. The output of the curve-fitting subroutine is the set of parameter values that optimize the objective function.

The first input to the subroutine is the choice of the parameters to be fit.
The curve-fitting subroutine fits three parameters.
One of these three (namely the mode-matching parameter $\gamma$ in Algorithm~\ref{Alg:Calibration} or the $|\theta_{ij}|$ or $\beta_{ii'jj'}$ value in Algorithm~\ref{Alg:PhaseCalcNChannel}) is related to the shape of the curve, whereas the other two are related to the ordinate scaling and the abscissa shift of the curve respectively.
The ordinate scaling factor comprises the unknown losses $\{\kappa_{i},\nu_{j}\}$, transmission factors $\{\lambda_{i},\mu_{j}\}$ and the incident photon-pair count. 
The horizontal shift factor accounts for the unknown zero of the time delay between the incident photons.
The algorithm returns the values of the shape parameter, the abscissa shift and the ordinate scaling that best fit the given coincidence curves.

The objective function quantifies the goodness of fit between the experimental data and the parameterized curve.
We use a weighted sum 
\begin{equation}
\sum_{\tau \in T} w(\tau)|C^\mathrm{exp}(\tau) - C'(\tau)|^2 
\end{equation}of squares between the experimental data and the fitted curve as the objective function~\cite{Strutz2010} for weighs $w(\tau)$.
We assume that the pdfs of the residues are proportional to $\sqrt{C^{\mathrm{exp}}(\tau)}$ and we assign the weights
\begin{equation}
 w(\tau) = 
 \begin{cases}
 {1}/{C^{\mathrm{exp}}(\tau)} & \text{if } C^{\mathrm{exp}}(\tau) \ne 0\\
 \quad 1 & \text{if } C^{\mathrm{exp}}(\tau) = 0
 \end{cases}
\end{equation}
to the squared sum of residues.
In case the pdf's of the residuals for different values of $\tau$ is not known, standard methods for non-parametric estimation of residual distribution can be employed to estimate the pdf's~\cite{Akritas2001,Chen2003}. 
Thus, the curve fitting algorithm returns those values of the fitting parameters that that minimize weighted sum of squared residues between experimental and fitted data. 

The curve-fitting procedure optimizes the fitness function over the domain of the fitting parameter values.
Like other optimization procedures, the convergence of curve fitting is sensitive to the initial guesses of the fitting parameters.
The following heuristics give good guesses for the three fitting parameters.
We guess the ordinate scaling as the ratio
\begin{equation}
\frac{C^{\mathrm{exp}}(\infty)}{C_{ii'jj'}(\infty)}
\end{equation}
 of the experimental coincidence counts
\begin{equation}
C^{\mathrm{exp}}(\infty) \defeq \frac{C^{\mathrm{exp}}(\tau_1) + C^{\mathrm{exp}}(\tau_\ell)}{2},
\end{equation}
to the coincidence probability $C_{ii'jj'}(\infty)$ for large (compared to the temporal length of the photon) time-delay values.
The $\gamma$ value is guessed for Algorithm~\ref{Alg:Calibration} as the ratio of the visibility of the experimental curve to the expected visibility in the curve. 
The initial guesses for $\vartheta \equiv |\theta_{ij}|$ and $\vartheta \equiv \beta_{ii'jj'}$ are based on the known estimate of $\gamma$ and the visibility 
\begin{equation}
 V = \frac{2\gamma\cos^2\vartheta\sin^2\vartheta}{\cos^4\vartheta + \sin^4\vartheta}.
\end{equation}
of the curve.
As there are four kinds of curves (see Figure~\ref{Figure:CoincidenceCurveShapes}) possible for different values of the shape parameter ($\gamma, |\theta_{ij}|, \beta_{ii'j'}$), another approach is to perform curve fitting four times, each time with a value from the set $\pi/4,3\pi/4,5\pi/4,7\pi/4$ of initial guesses and choose the fitted parameters that optimize the objective function. 
Finally, the initial value of the abscissa shift parameter is guessed such that the global maxima or minima (whichever is further from the mean of the coincidence-count values over $\tau$) of the coincidence curve is at zero time delay.

In summary, the curve fitting procedure uses the measured coincidence counts, the objective function and the initial guesses to compute the best fit parameters. 
This completes our description of the curve-fitting procedure and of heuristics that can be employed to computed the initial guesses for the fitted parameters.

\section*{References}

\bibliography{AccurateAndPrecise}

\end{document}